\documentclass[nologo]{jmi}
\usepackage{graphicx,amsmath}
\title{Self-adaptive moving mesh schemes for short pulse
type equations and their Lax pairs}
\author{BF}{Bao-Feng Feng}{feng@utpa.edu}
\author{KM}{Ken-ichi Maruno}{kmaruno@waseda.jp}
\author{YO}{Yasuhiro Ohta}{ohta@math.sci.kobe-u.ac.jp}
\affiliation{BF}{Department of Mathematics,
The University of Texas-Pan American,
Edinburg, Texas 78541, USA}
\affiliation{KM}{
Department of Mathematics,
The University of Texas-Pan American,
Edinburg, Texas 78541, USA\\
Present address: Department of Applied Mathematics, 
School of Fundamental Science and Engineering, Waseda University, 3-4-1 Okubo,
Shinjuku-ku, Tokyo 169-8555, Japan}%
\affiliation{YO}{Department of Mathematics,
Kobe University, Rokko, Kobe 657-8501, Japan}%
\abstract{%
Integrable self-adaptive moving mesh schemes for short pulse 
type equations (the short pulse equation, the coupled short pulse
equation, and the complex short pulse equation) are investigated. 
Two systematic methods, 
one is based on bilinear equations and another is based on Lax pairs, 
are shown.  
Self-adaptive moving mesh schemes consist of two semi-discrete
equations in which the time is continuous and the space is discrete. 
In self-adaptive moving mesh schemes, one of two equations is an evolution
equation of mesh intervals which is deeply related to 
a discrete analogue of a reciprocal (hodograph) transformation. 
An evolution equations of mesh intervals is 
a discrete analogue of a conservation law of an original equation, 
and a set of mesh intervals corresponds to a conserved density which
play an important role in generation of adaptive moving mesh. 
Lax pairs of self-adaptive moving mesh schemes 
for short pulse type equations are obtained by discretization 
of Lax pairs of short pulse type equations, thus the existence of Lax
pairs guarantees the integrability of self-adaptive moving mesh schemes 
for short pulse type equations.   
It is also shown that self-adaptive moving mesh schemes for short pulse type 
equations provide good numerical results by using standard time-marching methods
such as the improved Euler's method.}
\begin{document}
\maketitle

\section{Introduction}

The studies of discrete integrable systems were initiated in the middle of 
1970s. 
Hirota discretized various soliton equations such as the KdV, the mKdV,
and the sine-Gordon equations based on the bilinear
equations
\cite{Hirota:difference1,Hirota:difference2,Hirota:difference3,Hirota:difference4,Hirota:difference5},
Ablowitz and Ladik proposed a method of integrable
discretizations of soliton equations, including the nonlinear Schr\"odinger equation and the
modified KdV (mKdV) equation, based on the Ablowitz-Kaup-Newell-Segur
(AKNS) form\cite{Ablowitz-Ladik:1975,Ablowitz-Ladik:1976,Ablowitz-Ladik:1976-2,
Ablowitz-Ladik:1977,AS:book}. 
Following the pioneering works of Hirota and Ablowitz-Ladik, 
the studies of discrete integrable systems have been expanded in diverse areas 
(see, for example, \cite{Grammaticos:book,Suris:book,Bobenko:book,Bobenko:book2}).

It is known that there is a class of soliton equations which are derived
from the Wadati-Konno-Ichikawa (WKI) type $2\times 2$ linear 
system\cite{WKI1,WKI2,AS:book}. 
Soliton equations in the WKI class are 
transformed to certain soliton equations which are derived from 
the AKNS type $2\times 2$ linear system through reciprocal (hodograph) 
transformations\cite{Ishimori1,Ishimori2,Wadati-Sogo,Rogers-Schief,AS:book}. 

Integrable discretization of soliton equations in the
WKI class had been regarded as a difficult problem until recently. 
A systematic treatment of reciprocal (hodograph) transformations 
in integrable discretizations had been unknown for three decades. 
Recently, the present authors proposed integrable discretizations of some
soliton equations in the WKI class by using the bilinear
method, and it was confirmed that those integrable discrete equations 
work effectively on numerical computations 
of the above class of soliton equations as self-adaptive moving mesh 
schemes\cite{dCH,dCHcom,dHS,SPE_discrete1,SPE_discrete2,
dDym}. 
However, the method employed in our previous papers 
was rather technical, thus it is not easy to extract a fundamental
structure of discretizations to apply this method to a broader 
class of nonlinear wave equations including nonintegrable systems.
 
The aim of this article is to present two systematic methods (in
sophisticated forms), one is based on bilinear equations (this method is regarded as 
an extension of Hirota's discretization method) 
and another is based on Lax pairs (this method is regarded as an
extension of Ablowitz-Ladik's discretization method), 
to construct self-adaptive moving mesh schemes for soliton equations in the
WKI class. 
We demonstrate how to construct self-adaptive moving mesh
schemes for short pulse type equations 
whose Lax pairs are written in the WKI type form which 
is transformed into the Ablowitz-Kaup-Newell-Segur (AKNS) type 
form by reciprocal (hodograph) transformations. 
We clarify that moving mesh is generated by following discrete
conservation law and mesh intervals are nothing but discrete conserved
densities which is a key of self-adaptive moving mesh schemes.    
Lax pairs of self-adaptive moving mesh schemes 
for short pulse type equations are constructed by discretization 
of Lax pairs of short pulse type equations.
It is also shown that self-adaptive moving mesh schemes for short pulse type 
equations provide good numerical 
results by using standard time-marching methods such as the improved Euler's method.

The short pulse (SP) equation\cite{SP:SW,SP:CJSW,Sakovich1,Sakovich2,Matsuno_SPE,Matsuno_SPEreview} 
\begin{equation}
u_{xt}
=u+\frac{1}{6}(u^3)_{xx}\,,\label{SP}
\end{equation} 
is linked with the so-called coupled dispersionless (CD) system\cite{Konno-Oono1,Hirota-Tsujimoto,Konno-Oono2,Konno}
\begin{eqnarray}
&& \frac{\partial \rho}{\partial T}-\frac{\partial}{\partial X}
\left(-\frac{u^2}{2}\right)=0\,,\label{CD-1}\\
&& \frac{\partial^2 u}{\partial X \partial T}=\rho u\,,\label{CD-2}
\end{eqnarray}
through the reciprocal transformation (this is often called the
hodograph transformation in many literatures)
\begin{equation}
x=X_{0}+\int_{X_{0}}^{X}\rho(\tilde{X},T) d\tilde{X} \,,\quad t=T\,,\label{SP-reciprocal}
\end{equation}
where $X_{0}$ is a constant. 
The reciprocal (hodograph) transformation (\ref{SP-reciprocal}) yields 
\begin{eqnarray}
&&\frac{\partial}{\partial X}=\rho \frac{\partial}{\partial x}\,,\\
&&\frac{\partial}{\partial T}=\frac{\partial}{\partial
 t}-\frac{u^2}{2}\frac{\partial}{\partial x}\,.  
\end{eqnarray}
Note that the reciprocal (hodograph) transformation (\ref{SP-reciprocal}) originates
from the conservation law (\ref{CD-1}). 
Applying the reciprocal (hodograph) transformation (\ref{SP-reciprocal}) to
(\ref{CD-2}) yields
\begin{equation}
\frac{\partial}{\partial x} 
\left(\frac{\partial}{\partial
 t}-\frac{u^2}{2}\frac{\partial}{\partial x}\right)u=u\,.  
\end{equation}
This can be rewritten as eq.(\ref{SP}). Thus the SP equation is
equivalent to the CD system with the reciprocal (hodograph) transformation. 
As we mentioned in our previous paper, the reciprocal (hodograph) transformation
between the CD system and the SP equation is 
nothing but the transformation between 
the Lagrangian coordinate and the Eulerian coordinate\cite{SPE_discrete2}. 

The CD system (\ref{CD-1}) and (\ref{CD-2}) 
can be derived from the compatibility condition of the
following linear $2\times2$ system (Lax pair)\cite{Konno-Oono1}:
\begin{equation}
\frac{\partial \Psi}{\partial X}=U\Psi\,,\qquad 
\frac{\partial \Psi}{\partial T}=V\Psi \,,\label{lax-CD}
\end{equation}
where 
\begin{equation}
U=-{\rm i} \lambda 
\left(
\begin{array}{cc}
\rho & u_X \\
u_X  & -\rho 
\end{array}
\right)\,,\quad 
V=
\left(
\begin{array}{cc}
\frac{{\rm i}}{4\lambda} & -\frac{u}{2} \\
\frac{u}{2}  & - \frac{{\rm i}}{4\lambda}
\end{array}
\right)\,,\label{lax-CD-UV}
\end{equation}
and $\Psi$ is a two-component vector. 
By applying the reciprocal (hodograph) transformation (\ref{SP-reciprocal}) 
into the above linear $2\times2$ system (Lax pair) 
(\ref{lax-CD}) and (\ref{lax-CD-UV}), 
we obtain the linear $2\times2$ system (Lax pair) for the short pulse equation\cite{Sakovich1}: 
\begin{equation}
\frac{\partial \Psi}{\partial x}=\tilde{U}\Psi\,,
\qquad \frac{\partial \Psi}{\partial t}=\tilde{V}\Psi \,,\label{lax-SP}
\end{equation}
where 
\begin{eqnarray}
&&\tilde{U}=-{\rm i} \lambda 
\left(
\begin{array}{cc}
1 & u_x \\
u_x  & -1 
\end{array}
\right)\,,\\ 
&& \tilde{V}=
\left(
\begin{array}{cc}
\frac{{\rm i}}{4\lambda}-\frac{{\rm i}\lambda}{2}u^2 & -\frac{u}{2}- \frac{{\rm
 i}\lambda}{2}u^2u_x \\
\frac{u}{2}- \frac{{\rm
 i}\lambda}{2}u^2u_x   & -\frac{{\rm i}}{4\lambda}+\frac{{\rm i}\lambda}{2}u^2
\end{array}
\right)\,,
\end{eqnarray}
which can be rewritten as 
\begin{eqnarray}
&&\tilde{U}=\lambda 
\left(
\begin{array}{cc}
1 & u_x \\
u_x  & -1 
\end{array}
\right)\,,\label{lax-SP-U}\\
&&\tilde{V}=
\left(
\begin{array}{cc}
\frac{1}{4\lambda}+\frac{\lambda}{2}u^2 & -\frac{u}{2}+ \frac{\lambda}{2}u^2u_x \\
\frac{u}{2}+\frac{\lambda}{2}u^2u_x   & -\frac{1}{4\lambda}-\frac{\lambda}{2}u^2
\end{array}
\right)\,,\label{lax-SP-V}
\end{eqnarray}
by replacing $\lambda$ by ${\rm i}\lambda$. 
This is nothing but the Lax pair of the SP equation. 
Note that the Lax pair of the SP equation is of the WKI
type\cite{Sakovich1}. 
In general, soliton equations derived from WKI-type eigenvalue problems
are transformed into soliton equations 
derived from AKNS-type eigenvalue
problems by reciprocal (hodograph) 
transformations\cite{Ishimori1,Ishimori2,Wadati-Sogo,Rogers-Schief}. 

\section{A self-adaptive moving mesh scheme for the SP equation and its Lax pair}

The SP equation (\ref{SP}) can be discretized 
by means of the following two methods:\vspace{2mm}\\
{\bf Method 1: The discretization method using bilinear equations} 
\begin{itemize}
\item Step 1: Transform the SP equation (\ref{SP}) into the 
CD system (\ref{CD-1}) and (\ref{CD-2}) by the
reciprocal (hodograph) transformation (\ref{SP-reciprocal}). 
\item Step 2: Transform the CD system into the bilinear equations.
\item Step 3: Discretize the bilinear equations of the CD system. 
\item Step 4: Transform the (semi-)discrete bilinear equations into the (semi-)discrete CD
system.
\item Step 5: Discretize the reciprocal (hodograph) transformation and transform 
the (semi-)discrete CD system via the discrete reciprocal (hodograph) transformation. 
\end{itemize}
{\bf Method 2: the discretization method using a Lax pair}
\begin{itemize}
\item Step 1: Transform the Lax pair of the SP equation (\ref{SP}) by the
reciprocal (hodograph) transformation (\ref{SP-reciprocal}). The Lax pair obtained
      by the reciprocal (hodograph) transformation is the one of  into
      the Lax pair of the CD system (\ref{CD-1}) and (\ref{CD-2}). 
 
\item Step 2: Discretize the Lax pair of the CD system. The
      compatibility condition of the discretized Lax pair yields the
      (semi-)discrete CD system. 

\item Step 3: Discretize the reciprocal (hodograph) transformation and 
transform the discretized Lax pair of the (semi-)discrete CD system via the discrete
      reciprocal (hodograph) transformation. 

\item Step 4: The compatibility condition of the discretized Lax pair
      obtained in Step 3 yields the
      (semi-)discrete SP equation. 
\end{itemize}

Since the SP equation (\ref{SP}) is equivalent to the CD system 
(\ref{CD-1}) and (\ref{CD-2}) with the
reciprocal (hodograph) transformation (\ref{SP-reciprocal}), 
the semi-discrete CD system with the discrete reciprocal (hodograph) transformation 
is equivalent to the semi-discrete SP equation.

Here we show the details of procedures to construct the self-adaptive moving mesh scheme for 
the SP equation by means of the above two methods. \vspace{0.6mm}\\
\noindent
{\bf Method 1:}\\
Step 1: The SP equation (\ref{SP}) is transformed into 
the CD system (\ref{CD-1}) and (\ref{CD-2}) via 
the reciprocal (hodograph) transformation (\ref{SP-reciprocal}).\\
\noindent
Step 2: The CD system (\ref{CD-1}) and (\ref{CD-2}) 
can be transformed into the bilinear equations
\begin{eqnarray} 
&&D_T^2 f \cdot f =\frac{1}{2} g^2\,, \label{SPE_bilinear2-1}\\
&&D_XD_T f \cdot g =fg\,, \label{SPE_bilinear2-2}
\end{eqnarray}
via the dependent variable transformation
\begin{equation}
u=\frac{g}{f}\,, \quad \rho=1-2 (\ln f)_{XT} \,.\label{DVT}
\end{equation}
Here $D_X$ and $D_T$ are Hirota's $D$-operators defined as 
\begin{eqnarray*}
&&D_X^mf\cdot g=(\partial_{X}-\partial_{X'})^mf(X)g(X')|_{X'=X}\,.  
\end{eqnarray*}
\noindent
Step 3: Discretize the space variable $X$ in the bilinear equations (\ref{SPE_bilinear2-1}) and 
(\ref{SPE_bilinear2-2}).  
\begin{eqnarray} 
&&D_T^2 f_k \cdot f_k =\frac{1}{2} g_k^2\,, \label{d-SPE_bilinear2-1}\\
&&\frac{1}{a}D_T (f_{k+1} \cdot g_{k}-f_{k}\cdot g_{k+1})
 =\frac{1}{2}(f_{k+1}g_k+f_kg_{k+1})\,. \nonumber\\
&& \label{d-SPE_bilinear2-2}
\end{eqnarray}
\noindent
Step 4: Consider the dependent variable transformation
\begin{equation}
u_k=\frac{g_k}{f_k}\,, \quad \rho_k=1-\frac{2}{a} \left(\ln \frac{f_{k+1}}{f_k}\right)_{T} \,,
\end{equation}
which is a discrete analogue of (\ref{DVT}). 
Then the bilinear equations (\ref{d-SPE_bilinear2-1}) and
(\ref{d-SPE_bilinear2-2}) are transformed into 
\begin{eqnarray}
&&\partial_T \rho_k-\frac{\left(-\frac{u_{k+1}^2}{2}\right)-
\left(-\frac{u_k^2}{2}\right)}{a}=0\,,\label{semidiscrete-CD1} \\
&&\partial_T \left(\frac{u_{k+1}-u_k}{a}\right)=\rho_k \frac{u_{k+1}+u_{k}}{2}\,,\label{semidiscrete-CD2}
\end{eqnarray}
which is a semi-discrete analogue of the CD system. \\
\noindent
Step 5: 
Consider a discrete analogue of the reciprocal (hodograph) transformation
\begin{equation}
x_k={X}_0+\sum_{j=0}^{k-1}a\rho_j\,,\label{discrete-reciprocal-SP}
\end{equation} 
where $x_0=X_0$.
Now we introduce the mesh interval
\begin{equation}
\delta_k=x_{k+1}-x_k\,.
\end{equation}
Note that the mesh interval satisfies the relation 
\begin{equation}
\delta_k=a\rho_k\,,
\end{equation}
so we can rewrite equations (\ref{semidiscrete-CD1}) and
(\ref{semidiscrete-CD2}) with the discrete reciprocal (hodograph) transformation 
(\ref{discrete-reciprocal-SP}) into the self-adaptive moving mesh scheme
for the SP equation
\begin{eqnarray}
&&\partial_T \delta_k=\frac{-u_{k+1}^2+u_k^2}{2}\,,\label{semidiscrete-SP1}\\
&&\partial_T(u_{k+1}-u_k)=\delta_k \frac{u_{k+1}+u_{k}}{2}\,,\label{semidiscrete-SP2}
\end{eqnarray}
where $\delta_k$ is related to $x_k$ by $\delta_k=x_{k+1}-x_k$ which
originates from the discrete reciprocal (hodograph) transformation
\begin{equation}
x_k={X_0}+\sum_{j=0}^{k-1}\delta_j\,.\label{semidiscrete-SP-reciprocal}
\end{equation}
The set of points $\{(x_k,u_k)\}_{k=0,1,\cdots}$ provides a solution 
of the semi-discrete SP equation. Note that the above discrete reciprocal
(hodograph) transformation can be interpreted as the transformation between 
Eulerian description and Lagrangian description in 
a discretized space\cite{SPE_discrete2}. 

The discrete reciprocal (hodograph) transformation 
(\ref{semidiscrete-SP-reciprocal}) yields 
\begin{eqnarray}
\frac{\Delta}{\Delta X_k}&=&\frac{\Delta}{a}=
\frac{\Delta x_k}{a}\frac{\Delta}{\Delta x_k}=\rho_k\frac{\Delta}{\Delta x_k}
=\rho_k\frac{\Delta}{\delta_k}\,,\\
\frac{\partial}{\partial T}&=&\frac{\partial}{\partial t}
+\frac{\partial x_k}{\partial T}\frac{\partial}{\partial x_k}
=\frac{\partial}{\partial t}
+\sum_{j=0}^{k-1}\frac{\partial \delta_j}{\partial
T}\frac{\partial}{\partial x_k}
\nonumber\\
&=&\frac{\partial}{\partial t}
+\left(\sum_{j=0}^{k-1}\frac{-u_{j+1}^2+u_j^2}{2}\right)
\frac{\partial}{\partial x_k}\nonumber\\
&=&\frac{\partial}{\partial t}
+\left(\frac{-u_{k+1}^2+u_0^2}{2}\right)
\frac{\partial}{\partial x_k}\nonumber\\
&=&\frac{\partial}{\partial t}
+\left(\frac{-u_{k+1}^2}{2}\right)
\frac{\partial}{\partial x_k}\,, \quad \text{if}\,\, u_0=0\,,
\end{eqnarray}
where $\Delta$ is a difference operator defined as $\Delta f_k\equiv
f_{k+1}-f_k$. 
Applying this to eq.(\ref{semidiscrete-SP2}), we obtain 
\begin{eqnarray}
&&\frac{1}{\delta_k}\frac{\partial (u_{k+1}-u_k)}{\partial t}
-\frac{u_{k+1}^2}{2}
\frac{1}{\delta_k}\frac{\partial (u_{k+1}-u_k)}{\partial x_k}\nonumber\\
&&\hspace{2cm} =\frac{u_{k+1}+u_{k}}{2}\,.
\end{eqnarray}
In the continuous limit $\delta_k\to 0$, this leads to the SP equation
(\ref{SP}). 

We remark that eq.(\ref{semidiscrete-CD1}) describes the evolution of the
mesh interval $\delta_k$, and this equation is nothing but a discrete analogue
of the conservation law (\ref{CD-1}). This means that the mesh interval
$\delta_k$ is a conserved density of the self-adaptive moving mesh scheme. 
Thus the mesh interval $\delta_k$ is determined by the semi-discrete
conservation law. 
From the semi-discrete conservation law, one can find the following
property: 
If $\frac{-u_{k+1}^2+u_k^2}{2}<0$, i.e., the slope
between $u_k^2$ and $u_{k+1}^2$ is positive, then the mesh interval
$\delta_k$ becomes smaller.
If $\frac{-u_{k+1}^2+u_k^2}{2}>0$, i.e., the slope
between $u_k^2$ and $u_{k+1}^2$ is negative, then the mesh interval
$\delta_k$ becomes larger. 
Thus this scheme creates refined mesh grid for given data $\{x_k,u_k\}$
for $k=0,1,2,\cdots, N$, i.e., mesh grid is refined in which slopes are steep. \vspace{2mm}\\
\noindent
{\bf Method 2:}\\
Step 1: \\
The Lax pair of the SP equation is given by (\ref{lax-SP}) with
(\ref{lax-SP-U}) and (\ref{lax-SP-V}). This is transformed into
(\ref{lax-CD}) with (\ref{lax-CD-UV}) via the reciprocal (hodograph) 
transformation (\ref{SP-reciprocal}). \\
\noindent
Step 2: By discretizing the Lax pair (\ref{lax-CD}) with (\ref{lax-CD-UV}), 
we obtain the following linear $2\times2$ system (Lax pair):
\begin{equation}
\Psi_{k+1}=U_k\Psi_k\,,\qquad 
\frac{\partial \Psi_k}{\partial T}=V_k\Psi_k \,,\label{lax-discreteCD}
\end{equation}
where 
\begin{eqnarray}
&&U_k= 
\left(
\begin{array}{cc}
1-{\rm i} \lambda a \rho_k & -{\rm i} \lambda \frac{u_{k+1}-u_k}{a}\\
-{\rm i} \lambda \frac{u_{k+1}-u_k}{a}  & 1+{\rm i} \lambda a \rho_k 
\end{array}
\right)\,,\quad \\
&&V_k=
\left(
\begin{array}{cc}
\frac{{\rm i}}{4\lambda} & -\frac{u_k}{2} \\
\frac{u_k}{2}  & - \frac{{\rm i}}{4\lambda}
\end{array}
\right)\,,\label{lax-discreteCD-UV}
\end{eqnarray}
where $\Psi_k$ is a two-component vector. 
The compatibility condition yields the semi-discrete CD system
(\ref{semidiscrete-CD1}) and (\ref{semidiscrete-CD2}). \\
\noindent
Step 3: Consider a discrete analogue of the reciprocal (hodograph) 
transformation (\ref{discrete-reciprocal-SP}). 
Now we introduce the mesh interval $\delta_k=x_{k+1}-x_k$ which 
satisfies the relation $\delta_k=a\rho_k$,  
so one can rewrite $U_k$ and $V_k$ by using lattice intervals $\delta_k$ and
replacing $\lambda$ by ${\rm i}\lambda$: 
\begin{eqnarray}
&&U_k= 
\left(
\begin{array}{cc}
1+\lambda \delta_k & \lambda \frac{u_{k+1}-u_k}{a}\\
\lambda \frac{u_{k+1}-u_k}{a}  & 1-\lambda \delta_k 
\end{array}
\right)\,,\quad \\
&&V_k=
\left(
\begin{array}{cc}
\frac{1}{4\lambda} & -\frac{u_k}{2} \\
\frac{u_k}{2}  & - \frac{1}{4\lambda}
\end{array}
\right)\,.\label{lax-discreteSP-UV}
\end{eqnarray}
\noindent
Step 4: 
This Lax pair provides 
(\ref{semidiscrete-SP1}) and (\ref{semidiscrete-SP2}) which is nothing but 
the self-adaptive moving mesh scheme for the SP equation. \vspace{0.5mm}

\noindent
{\bf Numerical simulations:}\\ 
Here we show some examples of numerical simulations using the
self-adaptive moving mesh scheme (\ref{semidiscrete-SP1}) and
(\ref{semidiscrete-SP2}). 
As a time marching method, we use the improved Euler's method. 

Multi-soliton solutions of the SP equation are given by 
\begin{eqnarray}
&&u=\frac{g}{f}\,, \quad \rho=1-2(\ln f)_{XT}\,,\\
&&x=X_{0}+\int_{X_{0}}^{X}\rho(\tilde{X},T) d\tilde{X} \,, \quad
 t=T\,,\\
&& f=\left|
\begin{matrix}
\mathcal{A}_N &I_N \\
\noalign{\vskip5pt} -I_N & \mathcal{B}_N
\end{matrix}
\right|
=\left|I_N+\mathcal{A}_N\mathcal{B}_N\right|
\,,\\
&& g={\left|
\begin{matrix}
\mathcal{A}_N &I_N & {\bf e}_N^\top\\
\noalign{\vskip5pt}
-I_N & \mathcal{B}_N & {\bf 0}^\top\\
\noalign{\vskip5pt}
{\bf 0} & -{\bf a}_N & 0
\end{matrix}
\right|}\,,
\end{eqnarray}
where 
\begin{eqnarray*}
&&\mathcal{A}_N\\
&&=\small\left(
\begin{matrix}
\frac{{\rm e}^{\xi_1+\xi_1}}{4(1/p_1+1/p_1)}
&\frac{{\rm e}^{\xi_1+\xi_2}}{4(1/p_1+1/p_2)}
&\cdots
&\frac{{\rm e}^{\xi_1+\xi_N}}{4(1/p_1+1/p_N)}
\\
\frac{{\rm e}^{\xi_2+\xi_1}}{4(1/p_2+1/p_1)}
&\frac{{\rm e}^{\xi_2+\xi_2}}{4(1/p_2+1/p_2)}
&\cdots
&\frac{{\rm e}^{\xi_2+\xi_N}}{4(1/p_2+1/p_N)}
\\
\vdots&\vdots&\ddots&\vdots
\\
\frac{{\rm e}^{\xi_N+\xi_1}}{4(1/p_N+1/p_1)}
&\frac{{\rm e}^{\xi_N+\xi_2}}{4(1/p_N+1/p_2)}
&\cdots
&\frac{{\rm e}^{\xi_N+\xi_N}}{4(1/p_N+1/p_N)}
\end{matrix}
\right)\normalsize
\,,
\end{eqnarray*}
\[
\mathcal{B}_N= 
\left(
\begin{matrix}
\frac{a_1a_1}{1/p_1+1/p_1}
&\frac{a_1a_2}{1/p_1+1/p_2}&\cdots
&\frac{a_1a_N}{1/p_1+1/p_N}
\\
\frac{a_2a_1}{1/p_2+1/p_1}
&\frac{a_2a_2}{1/p_2+1/p_2}
&\cdots
&\frac{a_2a_N}{1/p_2+1/p_N}
\\
\vdots&\vdots&\ddots&\vdots
\\
\frac{a_Na_1}{1/p_N+1/p_1}
&\frac{a_Na_2}{1/p_N+1/p_2}&\cdots
&\frac{a_Na_N}{1/p_N+1/p_N}
\end{matrix}
\right)\,,
\]
and $I_N$ is the $N\times N$ identity matrix, 
${\bf a}^\top$ is the transpose of ${\bf a}$, 
\begin{eqnarray*}
&&{\bf a}_N=(a_1,a_2,\cdots ,a_N)\,,\quad 
{\bf e}_N=(e^{\xi_1},e^{\xi_2},\cdots ,e^{\xi_N}) \,,\\ 
&& {\bf 0}=(0,0,\cdots , 0)\,, 
\end{eqnarray*}
\[
\xi_i=p_iX+\frac{1}{p_i}T\,,
\qquad 1\le i\le N\,. 
\]

For example, the $\tau$-functions $f$ and $g$ of the 2-soliton solution are written as 
\begin{eqnarray}
&&f=1+\frac{a_1^2p_1^2}{16}e^{2\xi_1}+\frac{a_2^2p_2^2}{16}e^{2\xi_2}
+\frac{a_1a_2p_1^2p_2^2}{2(p_1+p_2)^2}e^{\xi_1+\xi_2}\nonumber\\
&&\hspace{1cm} +\frac{a_1^2a_2^2p_1^2p_2^2}{256}\left(\frac{p_1-p_2}{p_1+p_2}\right)^4e^{2\xi_1+2\xi_2}\,,\\
&& g=a_1e^{\xi_1}+a_2e^{\xi_2}
+\frac{a_1a_2^2p_2^2}{16}\left(\frac{p_1-p_2}{p_1+p_2}\right)^2e^{\xi_1+2\xi_2}\nonumber\\
&&\hspace{1cm}
+\frac{a_1^2a_2p_1^2}{16}\left(\frac{p_1-p_2}{p_1+p_2}\right)^2e^{2\xi_1+\xi_2}\,,
\end{eqnarray}
where 
\[
\xi_i=p_iX+\frac{1}{p_i}T\,,
\qquad i=1,2\,. 
\]

There are two types of 2-loop soliton solutions and a type of breather
solutions. :\\
(1) Interactions of two loop solitons if both $a_1$ and $a_2$ are positive, 
or if both $a_1$ and $a_2$ are negative. 
The wave numbers $p_1$ and $p_2$ are chosen real.\\
(2) Interactions of a loop soliton and an anti-loop soliton if $a_1$ and $a_2$ have opposite
signs. The wave numbers $p_1$ and $p_2$ are chosen real. \\
(3) Breather solutions if the wave numbers $p_1$ and $p_2$ are
chosen complex and satisfy $p_2=p_1^*$ and $a_2=a_1^*$ in the above $\tau$-functions. 

\begin{figure}
\includegraphics[clip,width=6.8cm]{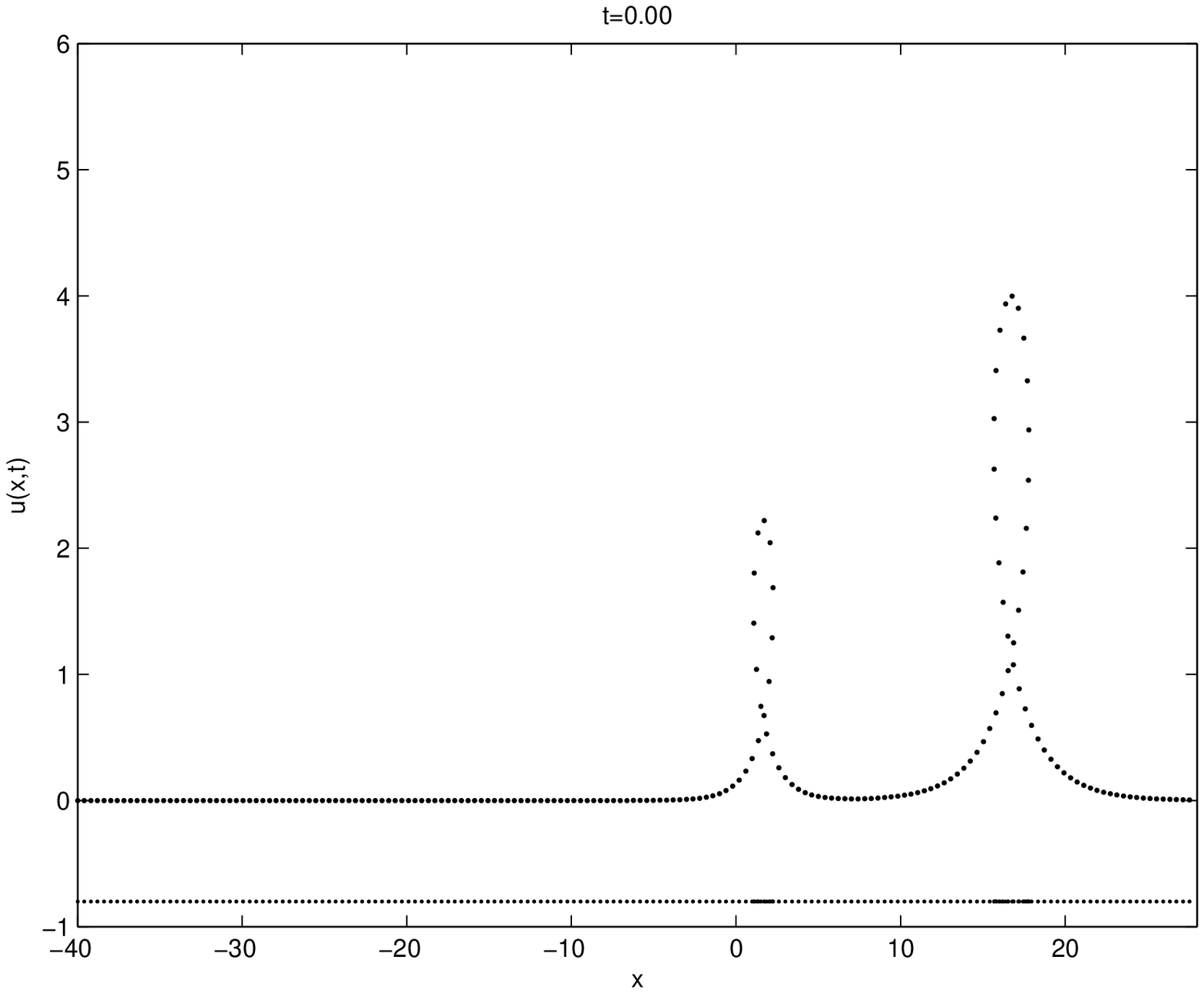} 
\includegraphics[clip,width=6.8cm]{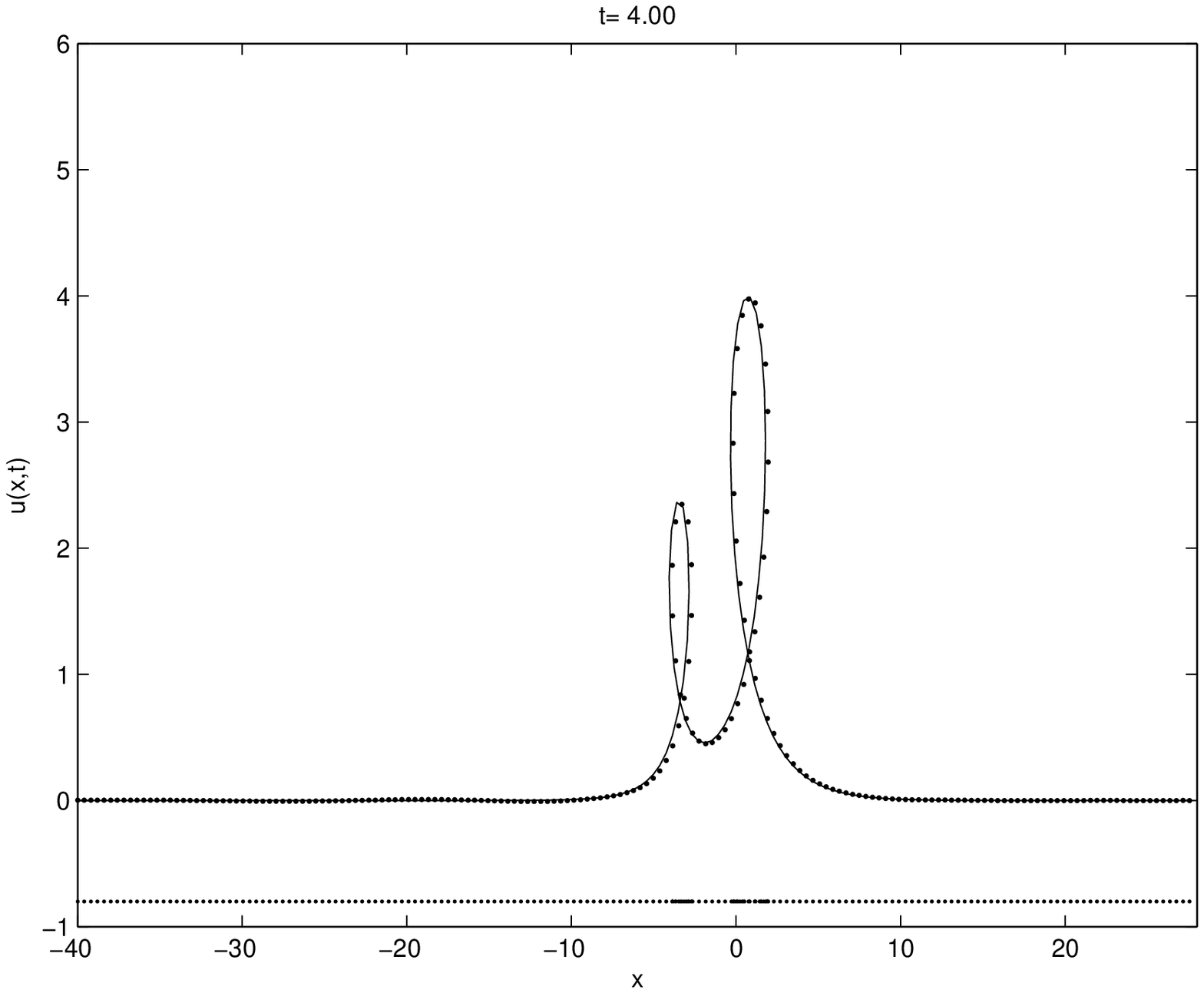} 
\includegraphics[clip,width=6.8cm]{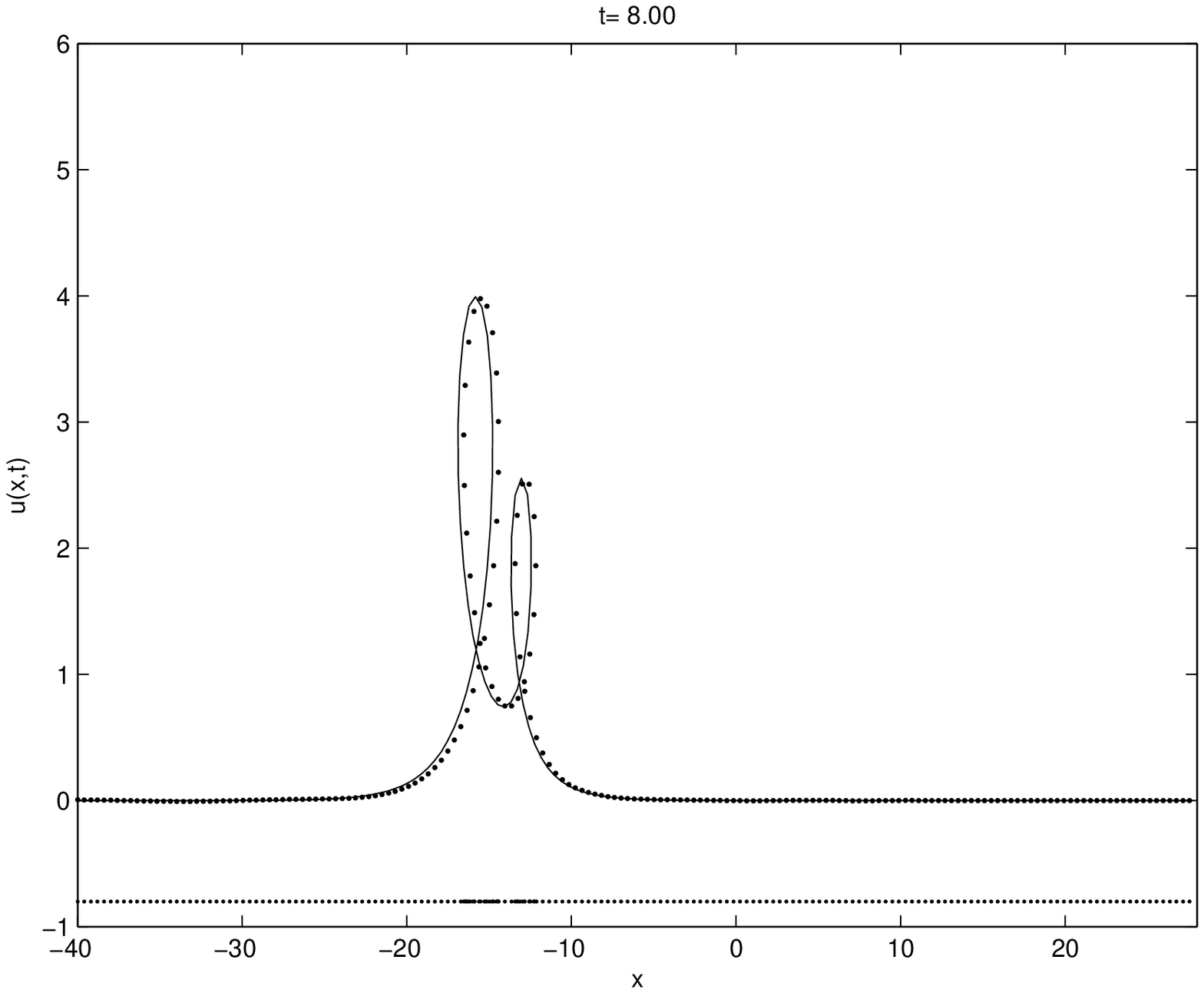} 
\includegraphics[clip,width=6.8cm]{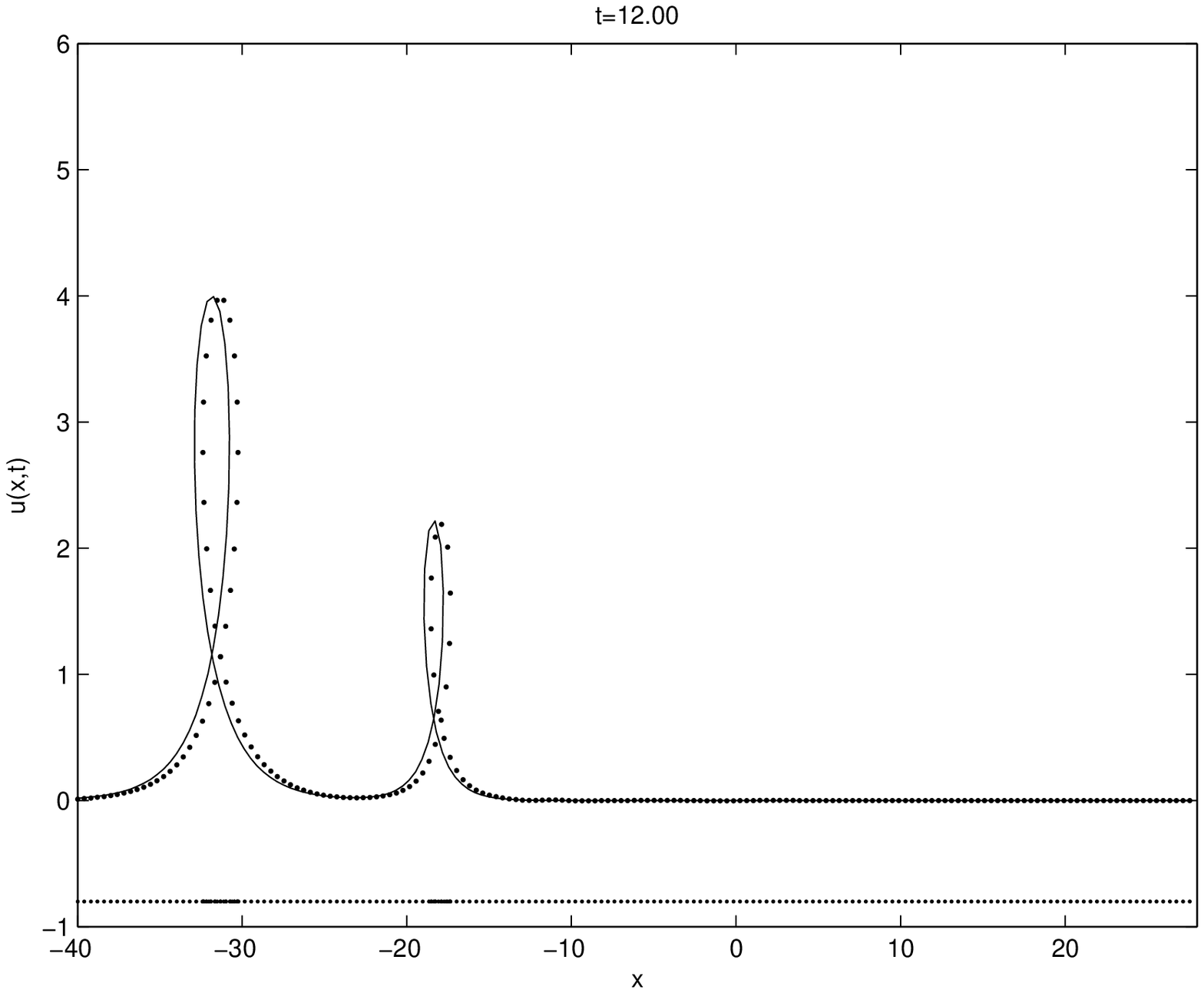} 
\caption{The numerical simulation of the 2-loop soliton solution of the SP
 equation. The points  
shows the numerical values, and the continuous curve shows the exact value, and
 the points on the bottom of graphs show the distribution of mesh grid
 points.  
$p_1=0.9$, $p_2=0.5$, $a_1=e^{-2}$, $a_2=e^{-8}$.
}
\label{2loop-numerics}
\end{figure}

\begin{figure}
\includegraphics[clip,width=6.8cm]{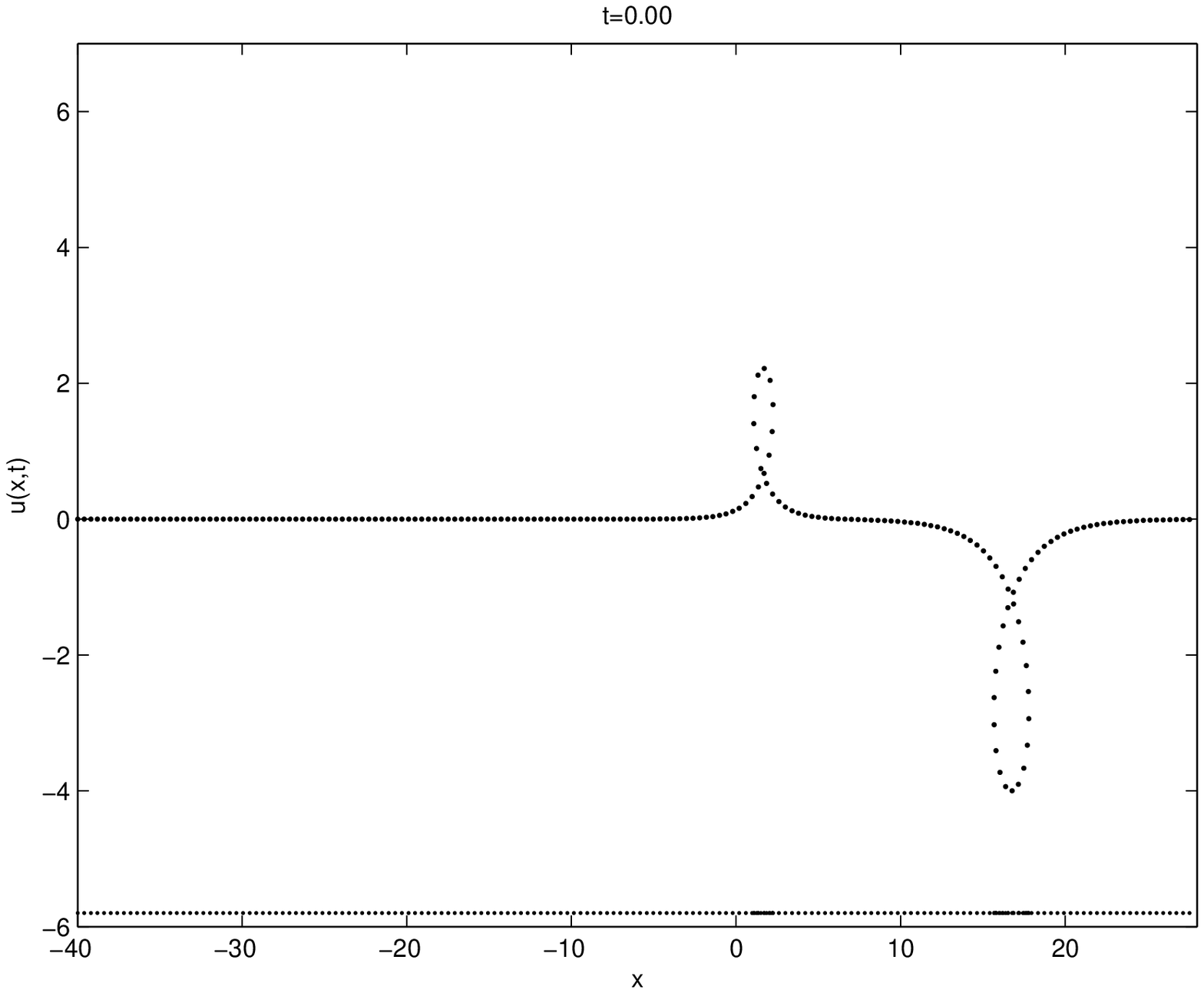} 
\includegraphics[clip,width=6.8cm]{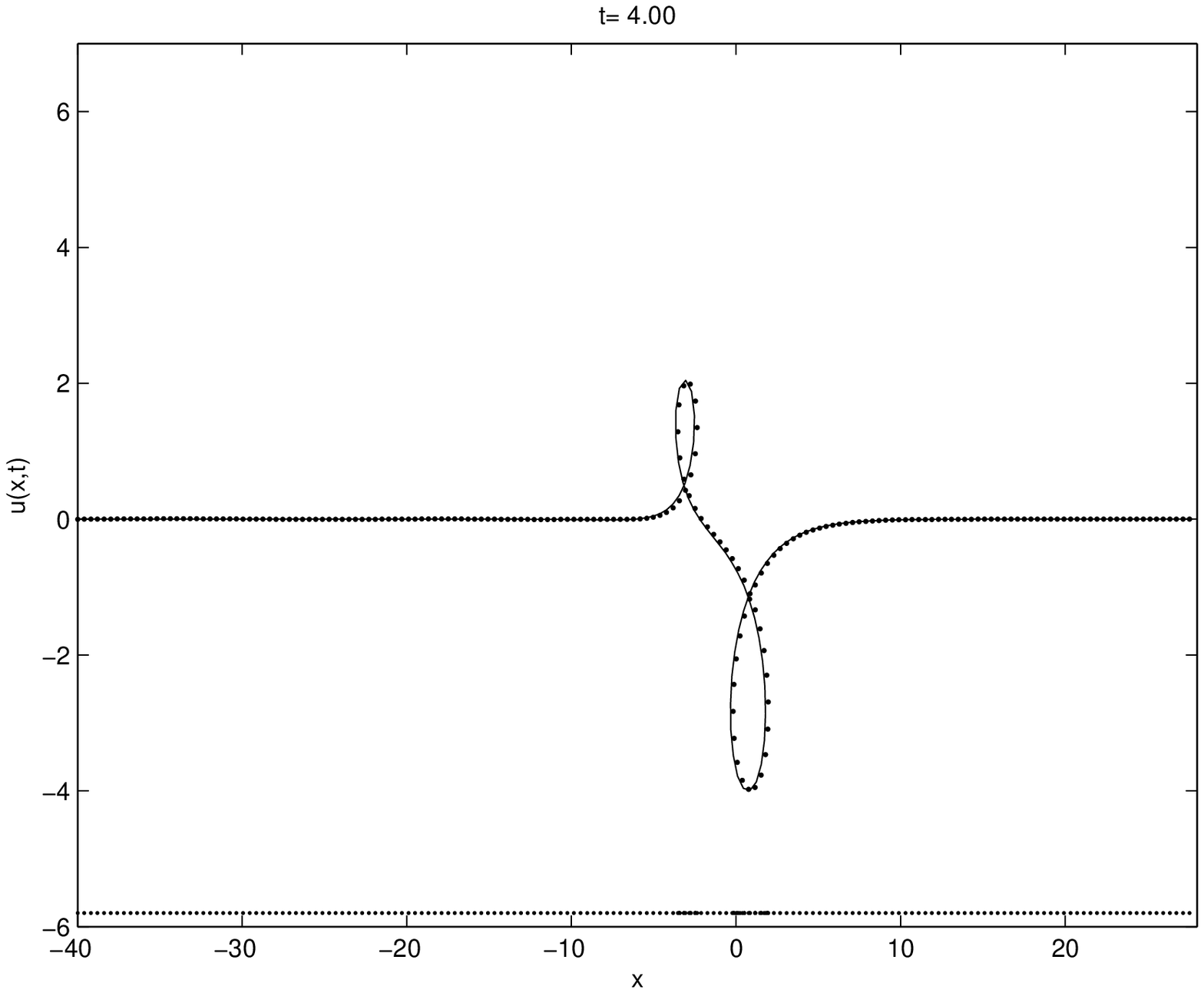} 
\includegraphics[clip,width=6.8cm]{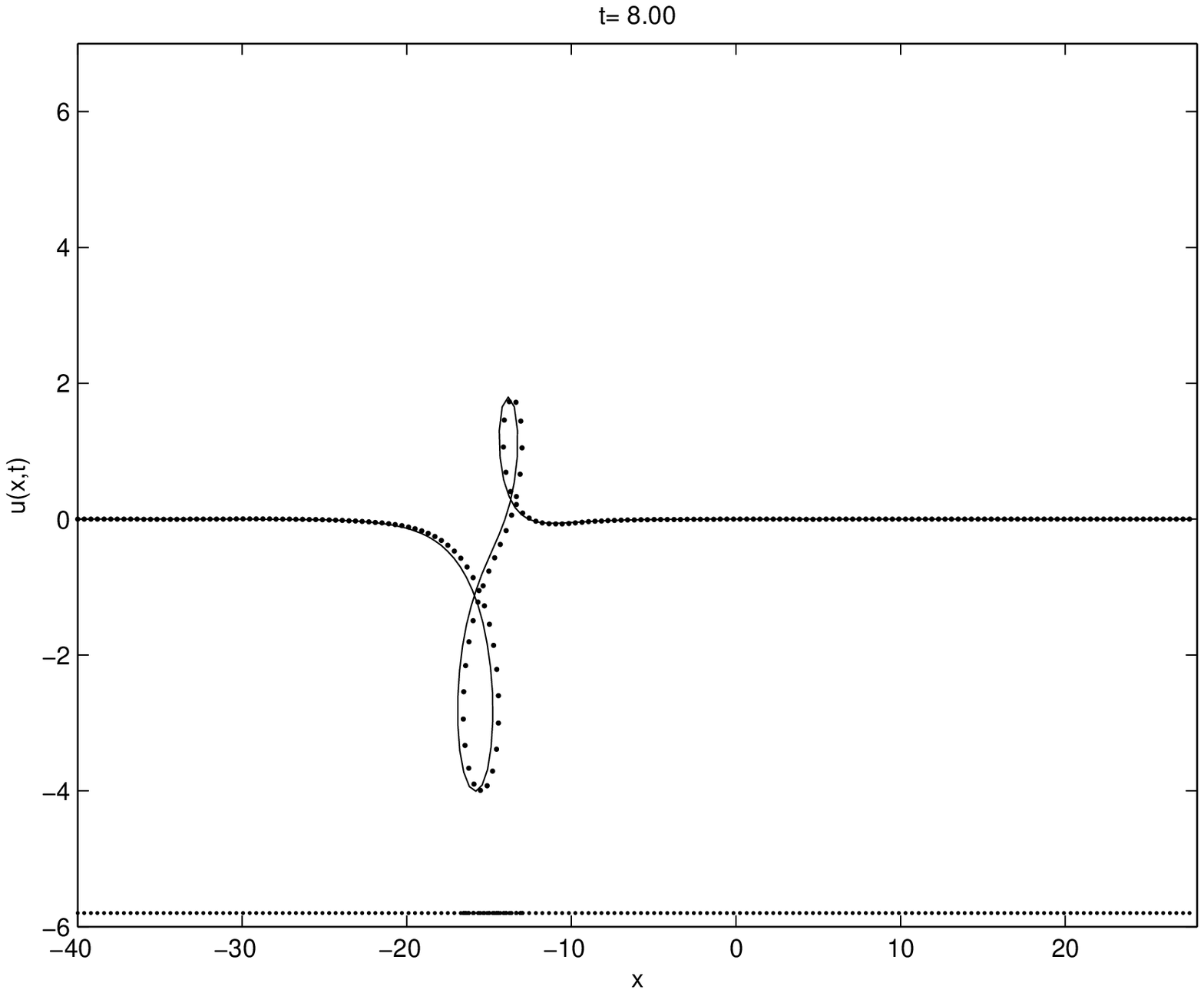} 
\includegraphics[clip,width=6.8cm]{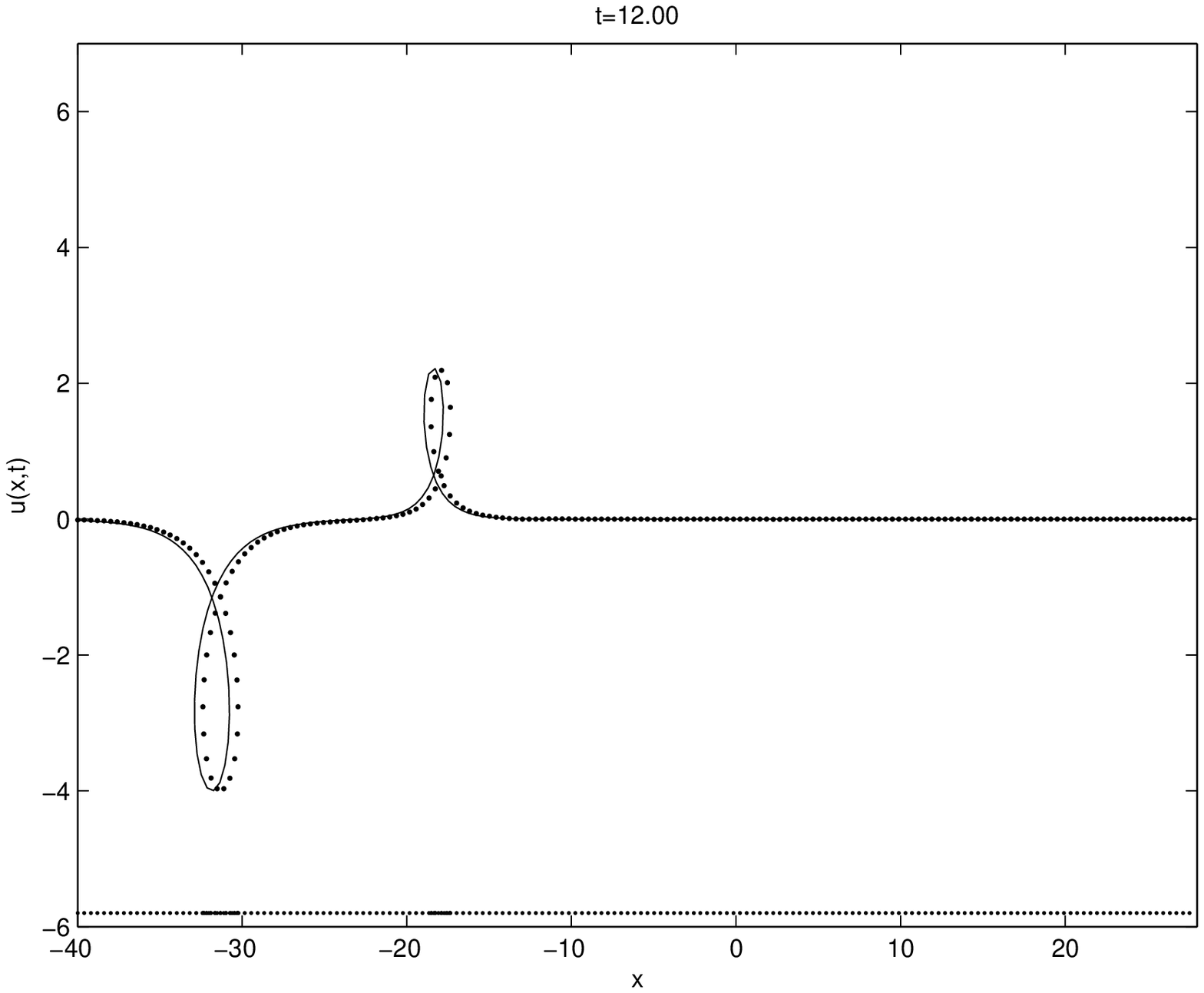} 
\caption{The numerical simulation of the loop-antiloop soliton solution of the SP
 equation. The points  
shows the numerical values, and the continuous curve shows the exact value, and
 the points on the bottom of graphs show the distribution of mesh grid
 points.  
$p_1=0.9$, $p_2=0.5$, $a_1=e^{-2}$, $a_2=-e^{-8}$.
}
\label{antiloop-numerics}
\end{figure}

\begin{figure}
\includegraphics[clip,width=6.8cm]{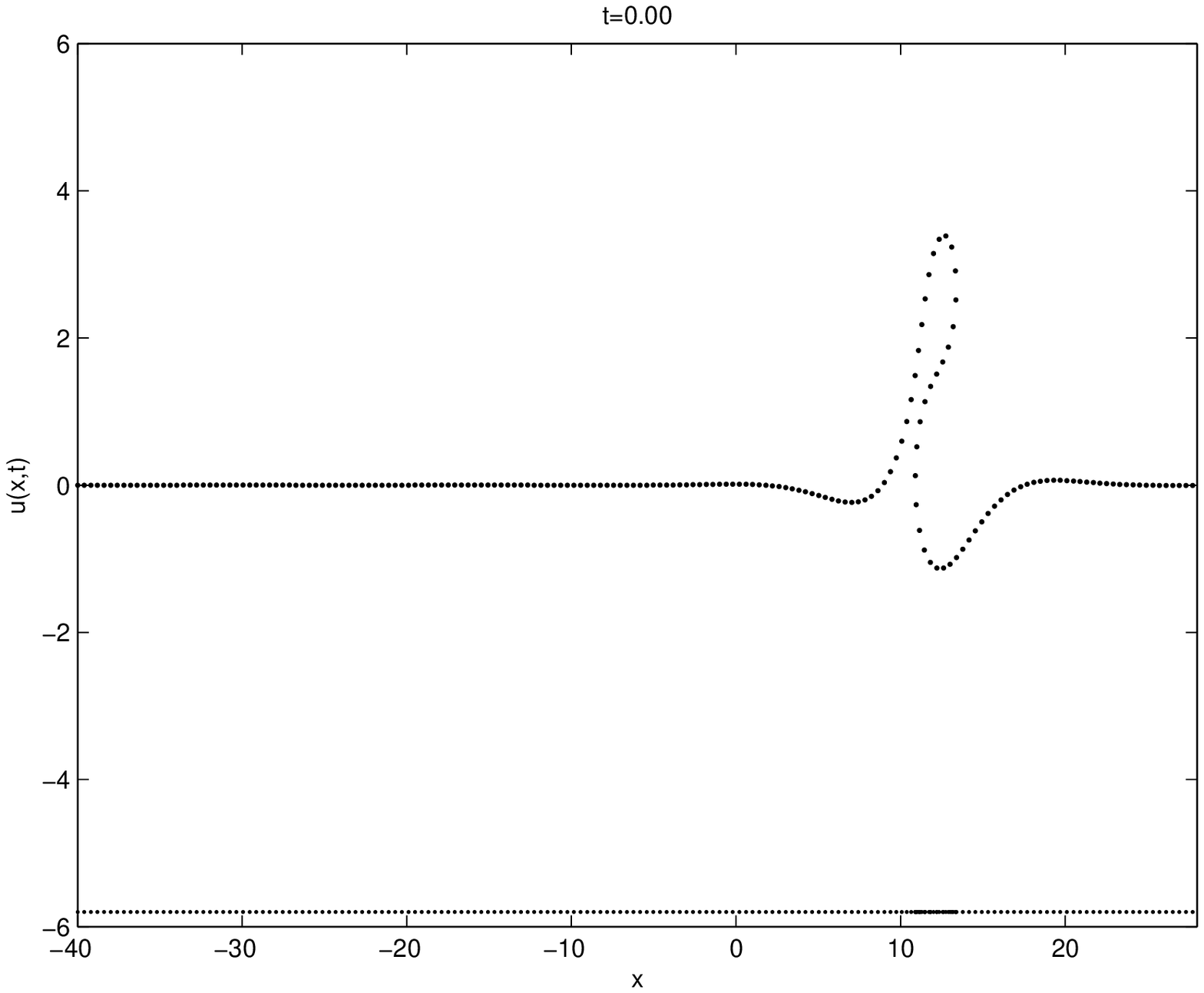} 
\includegraphics[clip,width=6.8cm]{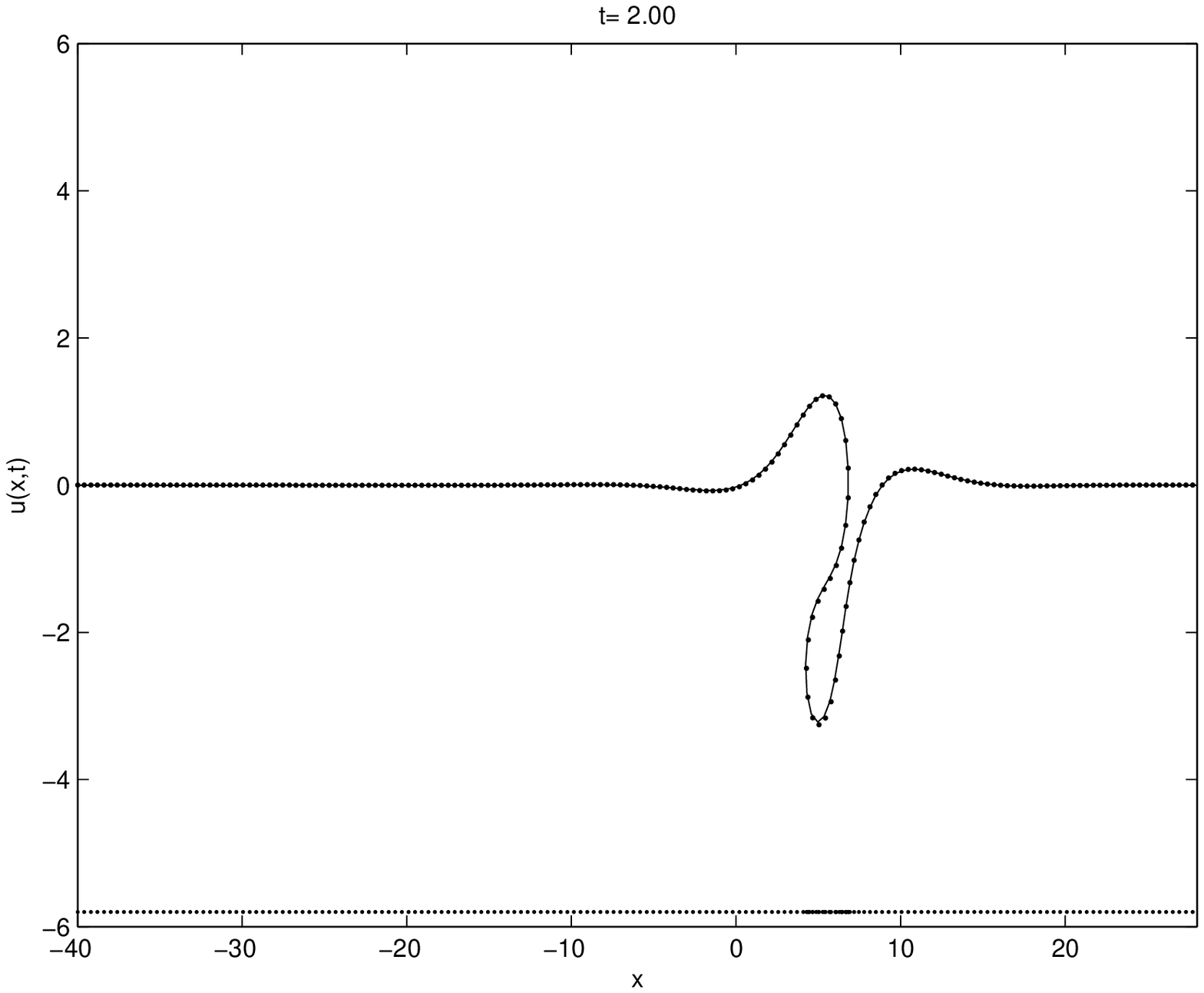} 
\includegraphics[clip,width=6.8cm]{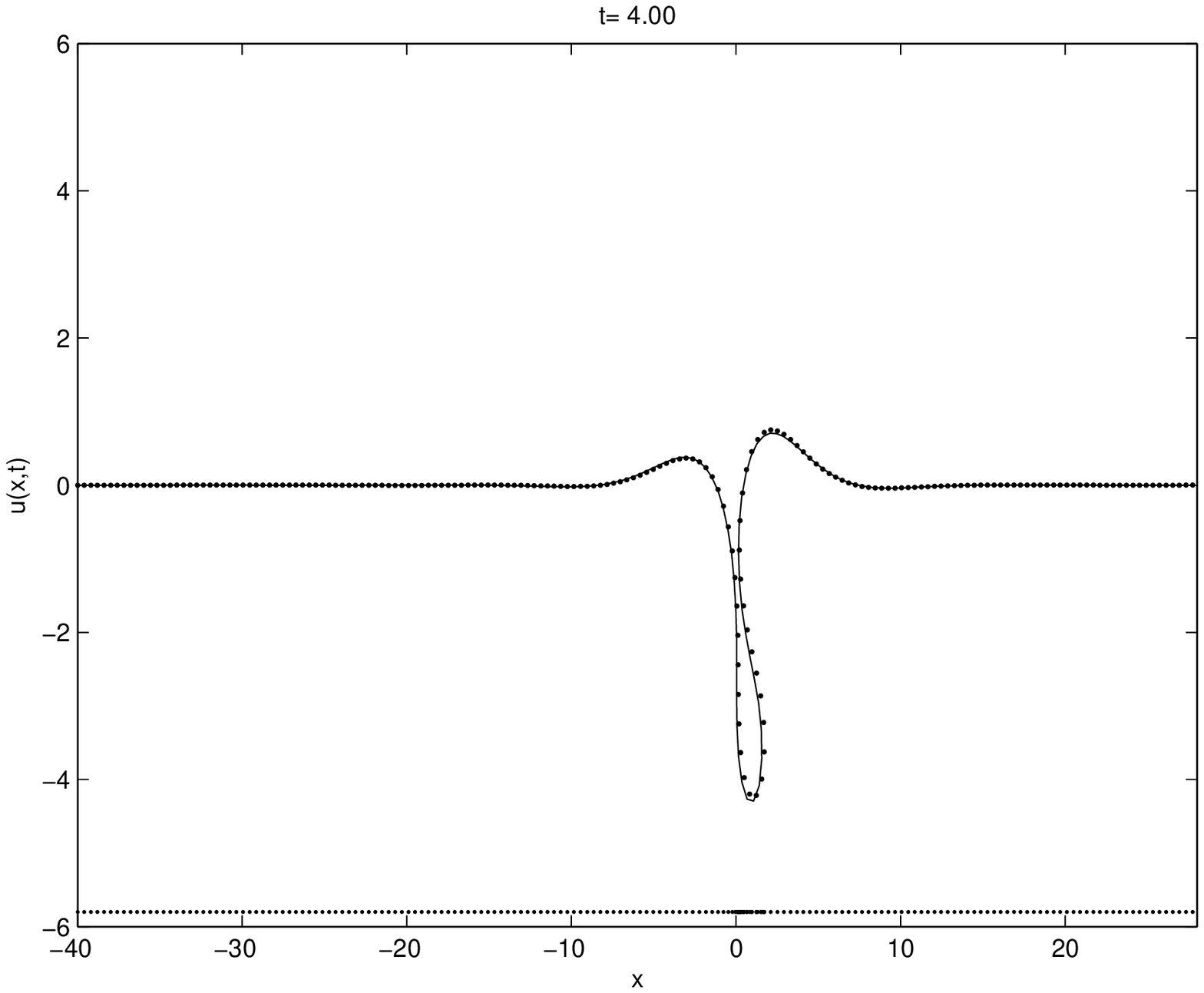} 
\includegraphics[clip,width=6.8cm]{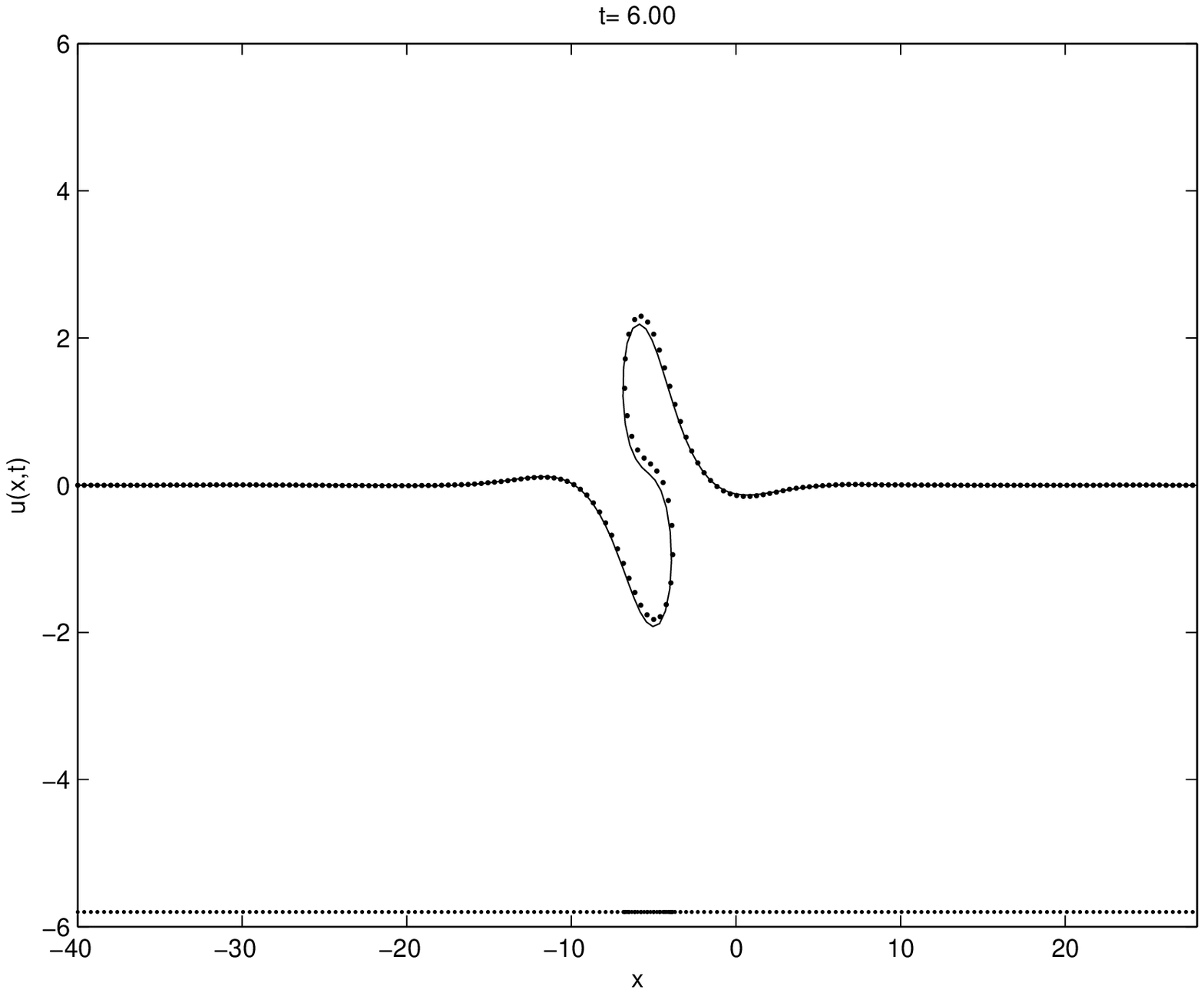} 
\caption{The numerical simulation of the breather solution of the SP
 equation. The points 
shows the numerical values, and the continuous curve shows the exact value, and
 the points on the bottom of graphs show the distribution of mesh grid
 points. $p_1=0.4+0.44{\rm i}$,
 $p_2=0.4-0.44{\rm i}$, $a_1=(1+{\rm i})e^{-2}$, $a_2=(1-{\rm i})e^{-8}$.}
\label{breather-numerics}
\end{figure}

Here we show three examples of numerical simulations of the SP equation. 
We use the number of mesh grid points $N=200$, the width of the computational
domain ${\rm D}=80$, and the time interval $dt=0.0001$.  
Figure \ref{2loop-numerics} shows the numerical simulation of the
2-loop soliton solution of the SP equation. 
Figure \ref{antiloop-numerics} shows the numerical simulation of 
the solution describing the interaction of a loop soliton and anti-loop soliton of the SP equation. 
In Figure \ref{breather-numerics}, we show the numerical simulation of the
breather solution of the SP equation. 
In these three examples, numerical results have good agreement with
exact solutions of the SP equation. It is possible to have more accurate numerical results
if we increase the number of mesh grid points and use smaller time steps. 

\section{Self-adaptive moving mesh schemes for the coupled short pulse
 equation and the complex short pulse equation}

By means of the above methods (Method 1 or Method 2) for constructing self-adaptive moving mesh
schemes, we can also construct self-adaptive moving mesh schemes for the
coupld SP equation and the complex SP equation. Here we show only the
results obtained by using Method 1 and Method 2. Note that both methods
give the same results.  

Consider the following linear $2\times2$ system:
\begin{equation}
\frac{\partial \Psi}{\partial X}=U\Psi\,,\qquad 
\frac{\partial \Psi}{\partial T}=V\Psi \,,\label{lax-coupledCD}
\end{equation}
where 
\begin{equation}
U=-{\rm i} \lambda 
\left(
\begin{array}{cc}
\rho & u_X \\
v_X  & -\rho 
\end{array}
\right)\,,\quad 
V=
\left(
\begin{array}{cc}
\frac{{\rm i}}{4\lambda} & -\frac{u}{2} \\
\frac{v}{2}  & - \frac{{\rm i}}{4\lambda}
\end{array}
\right)\,,\label{lax-coupledCD-UV}
\end{equation}
where $\Psi$ is a two-component vector. 
The compatibility condition yields the coupled CD system\cite{Kakuhata-Konno1,Kakuhata-Konno2} 
\begin{eqnarray}
&& \frac{\partial \rho}{\partial T}-\frac{\partial}{\partial X}
\left(-\frac{uv}{2}\right)=0\,,\label{coupledCD-1}\\
&& \frac{\partial^2 u}{\partial X \partial T}=\rho
 u\,,\label{coupledCD-2}\\
&& \frac{\partial^2 v}{\partial X \partial T}=\rho v\,.\label{coupledCD-3}
\end{eqnarray}
Note the coupled CD system (\ref{coupledCD-1}), (\ref{coupledCD-2}) and (\ref{coupledCD-3}) 
can be transformed into the bilinear equations
\begin{eqnarray} 
&&D_T^2 f \cdot f =\frac{1}{2} gh\,, \label{coupledCD_bilinear2-1}\\
&&D_XD_T f \cdot g =fg\,, \label{coupledCD_bilinear2-2}\\
&&D_XD_T f \cdot h =fh\,, \label{coupledCD_bilinear2-3}
\end{eqnarray}
via the dependent variable transformation
\begin{equation}
u=\frac{g}{f}\,, \quad v=\frac{h}{f}\,, \quad \rho=1-2 (\ln f)_{XT} \,.\label{DVT:coupledCD}
\end{equation}

Applying the reciprocal (hodograph) transformation (\ref{SP-reciprocal}) 
into the above linear problem 
(\ref{lax-coupledCD}) and (\ref{lax-coupledCD-UV}), 
we obtain the linear $2\times2$ system (Lax pair) for the coupled SP equation\cite{Matsuno-CSPE}: 
\begin{equation}
\frac{\partial \Psi}{\partial x}=\tilde{U}\Psi\,,
\qquad \frac{\partial \Psi}{\partial t}=\tilde{V}\Psi \,,
\end{equation}
where 
\begin{eqnarray}
&&\tilde{U}=-{\rm i} \lambda 
\left(
\begin{array}{cc}
1 & u_x \\
v_x  & -1 
\end{array}
\right)\,,\\ 
&& \tilde{V}=
\left(
\begin{array}{cc}
\frac{{\rm i}}{4\lambda}-\frac{{\rm i}\lambda}{2}uv & -\frac{u}{2}- \frac{{\rm
 i}\lambda}{2}uvu_x \\
\frac{v}{2}- \frac{{\rm
 i}\lambda}{2}uvv_x   & -\frac{{\rm i}}{4\lambda}+\frac{{\rm i}\lambda}{2}uv
\end{array}
\right)\,,
\end{eqnarray}
which can be rewritten as  
\begin{eqnarray}
&&\tilde{U}=\lambda 
\left(
\begin{array}{cc}
1 & u_x \\
v_x  & -1 
\end{array}
\right)\,,\\
&&\tilde{V}=
\left(
\begin{array}{cc}
\frac{1}{4\lambda}+\frac{\lambda}{2}uv & -\frac{u}{2}+ \frac{\lambda}{2}uvu_x \\
\frac{v}{2}+\frac{\lambda}{2}uvv_x   & -\frac{1}{4\lambda}-\frac{\lambda}{2}uv
\end{array}
\right)\,,
\end{eqnarray}
by replacing $\lambda$ by ${\rm i}\lambda$. 
The compatibility condition yields the coupled SP equation\cite{Folkert-CSPE,Matsuno-CSPE}
\begin{eqnarray}
&& u_{xt}=u+\frac{1}{2}(uvu_x)_x\,,\label{coupledSPeq1}\\
&& v_{xt}=v+\frac{1}{2}(uvv_x)_x\,.\label{coupledSPeq2}
\end{eqnarray}

Letting $u$ be a complex function and $v=u^*$ where $u^*$ is a complex conjugate of $u$, 
the compatibility condition of the above linear $2\times2$ systems 
yields the complex CD system\cite{Konno,Kotlyarov} 
\begin{eqnarray}
&& \frac{\partial \rho}{\partial T}-\frac{\partial}{\partial X}
\left(-\frac{|u|^2}{2}\right)=0\,,\label{complexCD-1}\\
&& \frac{\partial^2 u}{\partial X \partial T}=\rho
 u\,,\label{complexCD-2}\\
&& \frac{\partial^2 u^*}{\partial X \partial T}=\rho u^*\,,\label{complexCD-3}
\end{eqnarray}
and the complex SP equation\cite{Feng-CSPE}
\begin{eqnarray}
&& u_{xt}=u+\frac{1}{2}(|u|^2u_x)_x\,,\label{complexSPeq1}\\
&& u_{xt}^*=u^*+\frac{1}{2}(|u|^2u_x^*)_x\,.\label{complexSPeq2}
\end{eqnarray}
Note the complex CD system (\ref{complexCD-1}), (\ref{complexCD-2}) and (\ref{complexCD-3}) 
can be transformed into the bilinear equations
\begin{eqnarray} 
&&D_T^2 f \cdot f =\frac{1}{2} gg^*\,, \label{complexCD_bilinear2-1}\\
&&D_XD_T f \cdot g =fg\,, \label{complexCD_bilinear2-2}\\
&&D_XD_T f \cdot g^* =fg^*\,, \label{complexCD_bilinear2-3}
\end{eqnarray}
via the dependent variable transformation
\begin{equation}
u=\frac{g}{f}\,, \quad u^*=\frac{g^*}{f}\,, \quad \rho=1-2 (\ln f)_{XT} \,.\label{DVT:complexCD}
\end{equation}

Using the dependent variables $u^{({R})}$ and $u^{(\rm{I})}$ such that
$u^{({R})}={\rm Re} u$, $u^{(\rm{I})}={\rm Im} u$, the complex SP equation can be written
as 
\begin{eqnarray}
&& u_{xt}^{({R})}=u^{({R})}+\frac{1}{2}
\left(\left({u^{({R})}}^2+{u^{(\rm{I})}}^2\right)u_x^{({R})}\right)_x\,,\label{complexSPeq1-real}\\
&&
 u_{xt}^{(\rm{I})}=u^{(\rm{I})}+\frac{1}{2}
\left(\left({u^{({R})}}^2+{u^{(\rm{I})}}^2\right)u_x^{(\rm{I})}\right)_x
 \,.\label{complexSPeq2-imaginary}
\end{eqnarray}

Consider the following linear $2 \times 2$ system (\ref{lax-discreteCD})
with 
\begin{eqnarray}
&&U_k= 
\left(
\begin{array}{cc}
1-{\rm i} \lambda a \rho_k & -{\rm i} \lambda \frac{u_{k+1}-u_k}{a}\\
-{\rm i} \lambda \frac{v_{k+1}-v_k}{a}  & 1+{\rm i} \lambda a \rho_k 
\end{array}
\right)\,, \label{lax-discretecoupledCD-U}\\
&&V_k=
\left(
\begin{array}{cc}
\frac{{\rm i}}{4\lambda} & -\frac{u_k}{2} \\
\frac{v_k}{2}  & - \frac{{\rm i}}{4\lambda}
\end{array}
\right)\,,\label{lax-discretecoupledCD-V}
\end{eqnarray}
where $\Psi_k$ is a two-component vector. 
The compatibility condition 
(\ref{lax-discreteCD}) with (\ref{lax-discretecoupledCD-U}) 
and (\ref{lax-discretecoupledCD-V}) yields the semi-discrete coupled CD system 
\begin{eqnarray}
&&\partial_T \rho_k-\frac{\left(-\frac{u_{k+1}v_{k+1}}{2}\right)-
\left(-\frac{u_kv_k}{2}\right)}{a}=0\,,\label{semidiscrete-coupledCD1} \\
&&\partial_T \left(\frac{u_{k+1}-u_k}{a}\right)=\rho_k
 \frac{u_{k+1}+u_{k}}{2}\,,\label{semidiscrete-coupledCD2}\\
&&\partial_T \left(\frac{v_{k+1}-v_k}{a}\right)=\rho_k \frac{v_{k+1}+v_{k}}{2}\,.\label{semidiscrete-coupledCD3}
\end{eqnarray}
We can rewrite $U_k$ and $V_k$ by using lattice intervals $\delta_k(=a\rho_k=x_{k+1}-x_k)$ and
replacing $\lambda$ by ${\rm i}\lambda$: 
\begin{eqnarray}
&&U_k= 
\left(
\begin{array}{cc}
1+\lambda \delta_k & \lambda \frac{u_{k+1}-u_k}{a}\\
\lambda \frac{v_{k+1}-v_k}{a}  & 1-\lambda \delta_k 
\end{array}
\right)\,,\label{lax-discretecoupledSP-U}\\
&&V_k=
\left(
\begin{array}{cc}
\frac{1}{4\lambda} & -\frac{u_k}{2} \\
\frac{v_k}{2}  & - \frac{1}{4\lambda}
\end{array}
\right)\,.\label{lax-discretecoupledSP-V}
\end{eqnarray}
The compatibility condition of 
(\ref{lax-discreteCD}) with (\ref{lax-discretecoupledSP-U}) 
and (\ref{lax-discretecoupledSP-V}) 
provides the self-adaptive moving mesh scheme for the coupled SP
equation 
\begin{eqnarray}
&&\partial_T \delta_k=\frac{-u_{k+1}v_{k+1}+u_kv_k}{2}\,,\label{semidiscrete-coupledSP1}\\
&&\partial_T(u_{k+1}-u_k)=\delta_k
 \frac{u_{k+1}+u_{k}}{2}\,,\label{semidiscrete-coupledSP2}\\
&&\partial_T(v_{k+1}-v_k)=\delta_k \frac{v_{k+1}+v_{k}}{2}\,,\label{semidiscrete-coupledSP3}
\end{eqnarray}
where $x_k=X_k+\sum_{j=0}^{k-1}\delta_j$ and $\delta_k=x_{k+1}-x_k$,
$x_0=X_0$.

The discrete reciprocal (hodograph) transformation
$x_k=X_k+\sum_{j=0}^{k-1}\delta_j$ yields 
\begin{eqnarray}
\frac{\Delta}{\Delta X_k}&=&\frac{\Delta}{a}=
\frac{\Delta x_k}{a}\frac{\Delta}{\Delta x_k}=\rho_k\frac{\Delta}{\Delta x_k}
=\rho_k\frac{\Delta}{\delta_k}\,,\\
\frac{\partial}{\partial T}&=&\frac{\partial}{\partial t}
+\frac{\partial x_k}{\partial T}\frac{\partial}{\partial x_k}
=\frac{\partial}{\partial t}
+\sum_{j=0}^{k-1}\frac{\partial \delta_j}{\partial
T}\frac{\partial}{\partial x_k}
\nonumber\\
&=&\frac{\partial}{\partial t}
+\left(\sum_{j=0}^{k-1}\frac{-u_{k+1}v_{k+1}+u_kv_k}{2}\right)
\frac{\partial}{\partial x_k}\nonumber\\
&=&\frac{\partial}{\partial t}
+\left(\frac{-u_{k+1}v_{k+1}+u_0v_0}{2}\right)
\frac{\partial}{\partial x_k}\nonumber\\
&=&\frac{\partial}{\partial t}
+\left(\frac{-u_{k+1}v_{k+1}}{2}\right)
\frac{\partial}{\partial x_k}\,, \\
&\quad& \hspace{2.5cm} \text{if}\,\, u_0=0\,, v_0=0\,.\nonumber
\end{eqnarray}
Applying this to eqs.(\ref{semidiscrete-coupledSP2}) and
(\ref{semidiscrete-coupledSP3}), 
we obtain 
\begin{eqnarray}
&&\frac{1}{\delta_k}\frac{\partial (u_{k+1}-u_k)}{\partial t}
-\frac{u_{k+1}v_{k+1}}{2}
\frac{1}{\delta_k}\frac{\partial (u_{k+1}-u_k)}{\partial x_k}\nonumber\\
&&\hspace{2cm} =\frac{u_{k+1}+u_{k}}{2}\,,\\
&&\frac{1}{\delta_k}\frac{\partial (v_{k+1}-v_k)}{\partial t}
-\frac{u_{k+1}v_{k+1}}{2}
\frac{1}{\delta_k}\frac{\partial (v_{k+1}-v_k)}{\partial x_k}\nonumber\\
&&\hspace{2cm} =\frac{v_{k+1}+v_{k}}{2}\,.
\end{eqnarray}
In the continuous limit $\delta_k\to 0$, this leads to the coupled SP equation
(\ref{coupledSPeq1}) and (\ref{coupledSPeq2}). 

Note that the semi-discrete coupled CD system and the self-adaptive
moving mesh scheme for the coupled SP equation can be transformed into
the bilinear equations
\begin{eqnarray} 
&&D_T^2 f_k \cdot f_k =\frac{1}{2} g_kh_k\,, \label{d-coupledCD_bilinear2-1}\\
&&\frac{1}{a}D_T (f_{k+1} \cdot g_{k}-f_{k}\cdot g_{k+1})
 =\frac{1}{2}(f_{k+1}g_k+f_kg_{k+1})\,, \nonumber\\
&& \label{d-coupledCD_bilinear2-2}\\
&&\frac{1}{a}D_T (f_{k+1} \cdot h_{k}-f_{k}\cdot h_{k+1})
 =\frac{1}{2}(f_{k+1}h_k+f_kh_{k+1})\,, \nonumber\\
&& \label{d-coupledCD_bilinear2-3}
\end{eqnarray}
via the dependent variable transformation
\begin{eqnarray}
&&u_k=\frac{g_k}{f_k}\,, \quad v_k=\frac{h_k}{f_k}\,, \nonumber\\ 
&&\rho_k=\frac{\delta_k}{a}=1-\frac{2}{a} \left(\ln \frac{f_{k+1}}{f_k}\right)_{T} \,.
\end{eqnarray}

By letting $u_k$ be a complex function and adding a constraint $v_k=u_k^*$, 
we obtain the semi-discrete complex CD system
\begin{eqnarray}
&&\partial_T \rho_k-\frac{\left(-\frac{u_{k+1}v_{k+1}}{2}\right)-
\left(-\frac{|u_k|^2}{2}\right)}{a}=0\,,\label{semidiscrete-complexCD1} \\
&&\partial_T \left(\frac{u_{k+1}-u_k}{a}\right)=\rho_k
 \frac{u_{k+1}+u_{k}}{2}\,,\label{semidiscrete-complexCD2}\\
&&\partial_T \left(\frac{u_{k+1}^*-u_k^*}{a}\right)=\rho_k
 \frac{u_{k+1}^*+u_{k}^*}{2}\,,
\label{semidiscrete-complexCD3}
\end{eqnarray}
and the self-adaptive moving mesh scheme for the complex SP equation 
\begin{eqnarray}
&&\partial_T \delta_k=\frac{-|u_{k+1}|^2+|u_k|^2}{2}\,,\label{semidiscrete-complexSP1}\\
&&\partial_T(u_{k+1}-u_k)=\delta_k
 \frac{u_{k+1}+u_{k}}{2}\,,\label{semidiscrete-complexSP2}\\
&&\partial_T(u_{k+1}^*-u_k^*)=\delta_k \frac{u_{k+1}^*+u_{k}^*}{2}\,.\label{semidiscrete-complexSP3}
\end{eqnarray}
Note that the semi-discrete complex CD system and the self-adaptive
moving mesh scheme for the complex SP equation can be transformed into
the bilinear equations
\begin{eqnarray} 
&&D_T^2 f_k \cdot f_k =\frac{1}{2} g_kg_k^*\,, \label{d-complexCD_bilinear2-1}\\
&&\frac{1}{a}D_T (f_{k+1} \cdot g_{k}-f_{k}\cdot g_{k+1})
 =\frac{1}{2}(f_{k+1}g_k+f_kg_{k+1})\,, \nonumber\\
&& \label{d-complexCD_bilinear2-2}\\
&&\frac{1}{a}D_T (f_{k+1} \cdot g_{k}^*-f_{k}\cdot g_{k+1}^*)
 =\frac{1}{2}(f_{k+1}g_k^*+f_kg_{k+1}^*)\,, \nonumber\\
&& \label{d-complexCD_bilinear2-3}
\end{eqnarray}
via the dependent variable transformation
\begin{eqnarray}
&&u_k=\frac{g_k}{f_k}\,, \quad u_k=\frac{g_k^*}{f_k}\,, \nonumber\\
&&\rho_k=\frac{\delta_k}{a}=1-\frac{2}{a} \left(\ln \frac{f_{k+1}}{f_k}\right)_{T} \,.
\end{eqnarray}

Using the dependent variables $u_k^{({R})}$ and $u_k^{(\rm{I})}$ such that
$u_k^{({R})}={\rm Re} u_k$, $u_k^{(\rm{I})}={\rm Im} u_k$, the complex SP equation can be written
as 
\begin{eqnarray}
&&\partial_T \delta_k\nonumber\\
&&\quad =-\frac{1}{2}\left(\left(u_{k+1}^{(R)}\right)^2+\left(u_{k+1}^{(I)}\right)^2\right)\nonumber\\
&& \qquad \quad 
+\frac{1}{2}\left(\left(u_{k+1}^{(R)}\right)^2+\left(u_{k+1}^{(I)}\right)^2\right)\,,
\label{semidiscrete-complexSP1-realimaginary}\\
&&\partial_T(u_{k+1}^{(R)}-u_k^{(R)})=\delta_k
 \frac{u_{k+1}^{(R)}+u_{k}^{(R)}}{2}\,,\label{semidiscrete-complexSP2-real}\\
&&\partial_T(u_{k+1}^{(I)}-u_k^{(I)})=\delta_k
\frac{u_{k+1}^{(I)}+u_{k}^{(I)}}{2}\,.\label{semidiscrete-complexSP3-imaginary}
\end{eqnarray}

We remark that the discretization of the generalized CD systems
were proposed by Vinet and Yu recently\cite{Vinet-Yu1}. 
Our results are consistent with their results.

\noindent
{\bf Numerical simulations:}\\ 
Here we show some examples of numerical simulations of the complex SP equation using the
self-adaptive moving mesh scheme (\ref{semidiscrete-complexSP1}), 
(\ref{semidiscrete-complexSP2}) and (\ref{semidiscrete-complexSP3}). 
As a time marching method, we use the improved Euler's method. 

The multi-soliton solutions of the complex SP equation
(\ref{complexSPeq1}) and (\ref{complexSPeq2}) are given by the
following formula: 
\begin{eqnarray}
&&u=\frac{g}{f}\,, \quad  u^*=\frac{g^*}{f}\,, \quad \rho=1-2(\ln f)_{XT}\,,\\
&&x=X_{0}+\int_{X_{0}}^{X}\rho(\tilde{X},T) d\tilde{X} \,, \quad
 t=T\,,\\
&& f=\left|
\begin{matrix}
\mathcal{A}_N &I_N \\
\noalign{\vskip5pt} -I_N & \mathcal{B}_N
\end{matrix}
\right|
=\left|I_N+\mathcal{A}_N\mathcal{B}_N\right|
\,,\\
&& g={\left|
\begin{matrix}
\mathcal{A}_N &I_N & {\bf e}_N^\top\\
\noalign{\vskip5pt}
-I_N & \mathcal{B}_N & {\bf 0}^\top\\
\noalign{\vskip5pt}
{\bf 0} & -{\bf a}_N & 0
\end{matrix}
\right|}\,,\\
&& g^*={\left|
\begin{matrix}
\mathcal{A}_N &I_N & {\bf 0}^\top\\
\noalign{\vskip5pt}
-I_N & \mathcal{B}_N & {{\bf a}_N^*}^\top\\
\noalign{\vskip5pt}
{\bf e}_N^* & {\bf 0} & 0
\end{matrix}
\right|}\,,
\end{eqnarray}
where 
\begin{eqnarray*}
&&\mathcal{A}_N=\\
&&\small \left(
\begin{matrix}
\frac{{\rm e}^{\xi_1+\xi_1^*}}{4(1/p_1+1/p_1^*)}
&\frac{{\rm e}^{\xi_1+\xi_2^*}}{4(1/p_1+1/p_2^*)}
&\cdots
&\frac{{\rm e}^{\xi_1+\xi_N^*}}{4(1/p_1+1/p_N^*)}
\\
\frac{{\rm e}^{\xi_2+\xi_1^*}}{4(1/p_2+1/p_1^*)}
&\frac{{\rm e}^{\xi_2+\xi_2^*}}{4(1/p_2+1/p_2^*)}
&\cdots
&\frac{{\rm e}^{\xi_2+\xi_N^*}}{4(1/p_2+1/p_N^*)}
\\
\vdots&\vdots&\ddots&\vdots
\\
\frac{{\rm e}^{\xi_N+\xi_1^*}}{4(1/p_N+1/p_1^*)}
&\frac{{\rm e}^{\xi_N+\xi_2^*}}{4(1/p_N+1/p_2^*)}
&\cdots
&\frac{{\rm e}^{\xi_N+\xi_N^*}}{4(1/p_N+1/p_N^*)}
\end{matrix}
\right)\normalsize
\,,
\end{eqnarray*}
\[
\mathcal{B}_N= 
\left(
\begin{matrix}
\frac{a_1a_1^*}{1/p_1+1/p_1^*}
&\frac{a_2a_1^*}{1/p_2+1/p_1^*}&\cdots
&\frac{a_Na_1^*}{1/p_N+1/p_1^*}
\\
\frac{a_1a_2^*}{1/p_1+1/p_2^*}
&\frac{a_2a_2^*}{1/p_2+1/p_2^*}
&\cdots
&\frac{a_Na_2^*}{1/p_N+1/p_2^*}
\\
\vdots&\vdots&\ddots&\vdots
\\
\frac{a_1a_N^*}{1/p_1+1/p_N^*}
&\frac{a_2a_N^*}{1/p_2+1/p_N^*}&\cdots
&\frac{a_Na_N^*}{1/p_N+1/p_N^*}
\end{matrix}
\right)\,,
\]
and $I_N$ is the $N\times N$ identity matrix, 
${\bf a}^\top$ is the transpose of ${\bf a}$, 
\begin{eqnarray*}
&&{\bf a}_N=(a_1,a_2,\cdots ,a_N)\,,\quad 
{\bf e}_N=(e^{\xi_1},e^{\xi_2},\cdots ,e^{\xi_N}) \,,\\ 
&&{\bf a}_N^*=(a_1^*,a_2^*,\cdots ,a_N^*)\,,\quad 
{\bf e}_N^*=(e^{\xi_1^*},e^{\xi_2^*},\cdots ,e^{\xi_N^*}) \,,\\ 
&& {\bf 0}=(0,0,\cdots , 0)\,, 
\end{eqnarray*}
\[
\xi_i=p_iX+\frac{1}{p_i}T\,,
\quad 
\xi_i^*=p_i^*X+\frac{1}{p_i^*}T
\,,
\qquad 1\le i\le N\,. 
\]
Note that this formula is obtained from the gram type determinant 
solution of the coupled SP equation in Appendix. 

For example, the $\tau$-functions $f$, $g$ and $g^*$ of the 2-soliton
solution of the complex SP equation are written as 
\begin{eqnarray}
&&f=1+
\frac{a_1a_1^*p_1^2{p_1^*}^2}{4(p_1+p_1^*)^2}e^{\xi_1+\xi_1^*}
+\frac{a_1a_2^*p_1^2{p_2^*}^2}{4(p_1+p_2^*)^2}e^{\xi_1+\xi_2^*}\nonumber
\\
&&\quad +\frac{a_2a_1^*p_2^2{p_1^*}^2}{4(p_2+p_1^*)^2}e^{\xi_2+\xi_1^*}
+\frac{a_2a_2^*p_2^2{p_2^*}^2}{4(p_2+p_2^*)^2}e^{\xi_2+\xi_2^*}
\nonumber\\
&&\quad  
+\frac{a_1a_1^*a_2a_2^*p_1^2{p_1^*}^2p_2^2{p_2^*}^2(p_1-p_2)^2(p_1^*-p_2^*)^2}
{16(p_1+p_1^*)^2(p_1+p_2^*)^2(p_2+p_1^*)^2(p_2+p_2^*)^2}\nonumber \\
&& \qquad \times e^{\xi_1+\xi_2+\xi_1^*+\xi_2^*}\,,\\
&& g=a_1e^{\xi_1}+a_2e^{\xi_2}
+\frac{a_1a_2a_2^*(p_1-p_2)^2{p_2^*}^4}{4(p_1+p_2^*)^2(p_2+p_2^*)^2}e^{\xi_1+\xi_2+\xi_2^*}\nonumber\\
&&\hspace{1cm}
+\frac{a_1a_1^*a_2(p_1-p_2)^2{p_1^*}^4}{4(p_1+p_1^*)^2(p_2+p_1^*)^2}e^{\xi_1+\xi_1^*+\xi_2}\,,\\
&& g^*=a_1^*e^{\xi_1^*}+a_2^*e^{\xi_2^*}
+\frac{a_1^*a_2a_2^*(p_1^*-p_2^*)^2{p_2}^4}{4(p_1^*+p_2)^2(p_2+p_2^*)^2}e^{\xi_1^*+\xi_2+\xi_2^*}\nonumber\\
&&\hspace{1cm}
+\frac{a_1a_1^*a_2^*(p_1^*-p_2^*)^2{p_1}^4}{4(p_1^*+p_1)^2(p_1+p_2^*)^2}e^{\xi_1+\xi_1^*+\xi_2^*}\,,
\end{eqnarray}
where 
\[
\xi_i=p_iX+\frac{1}{p_i}T\,,
\qquad i=1,2\,. 
\]

\begin{figure}
\includegraphics[clip,width=6.8cm]{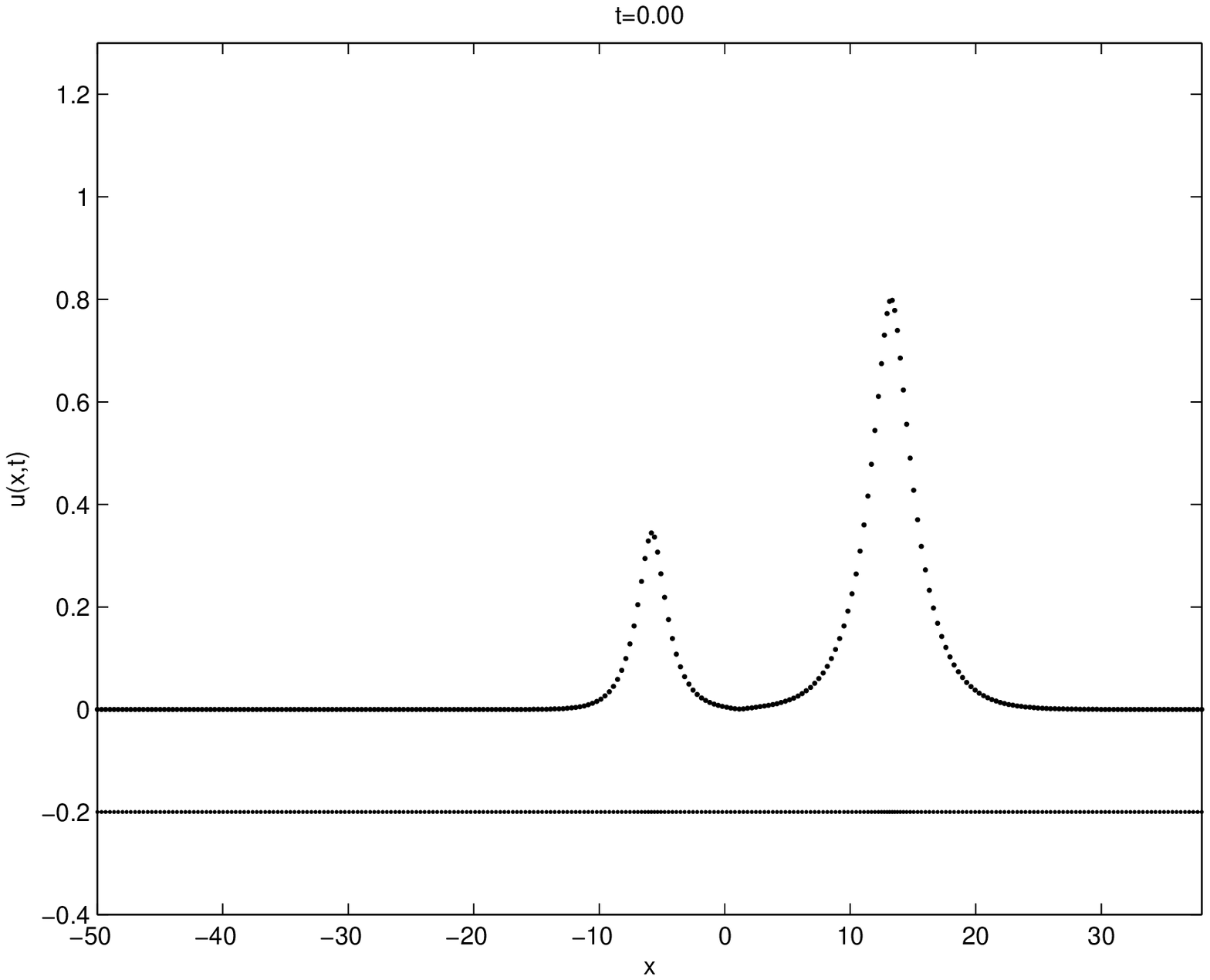} 
\includegraphics[clip,width=6.8cm]{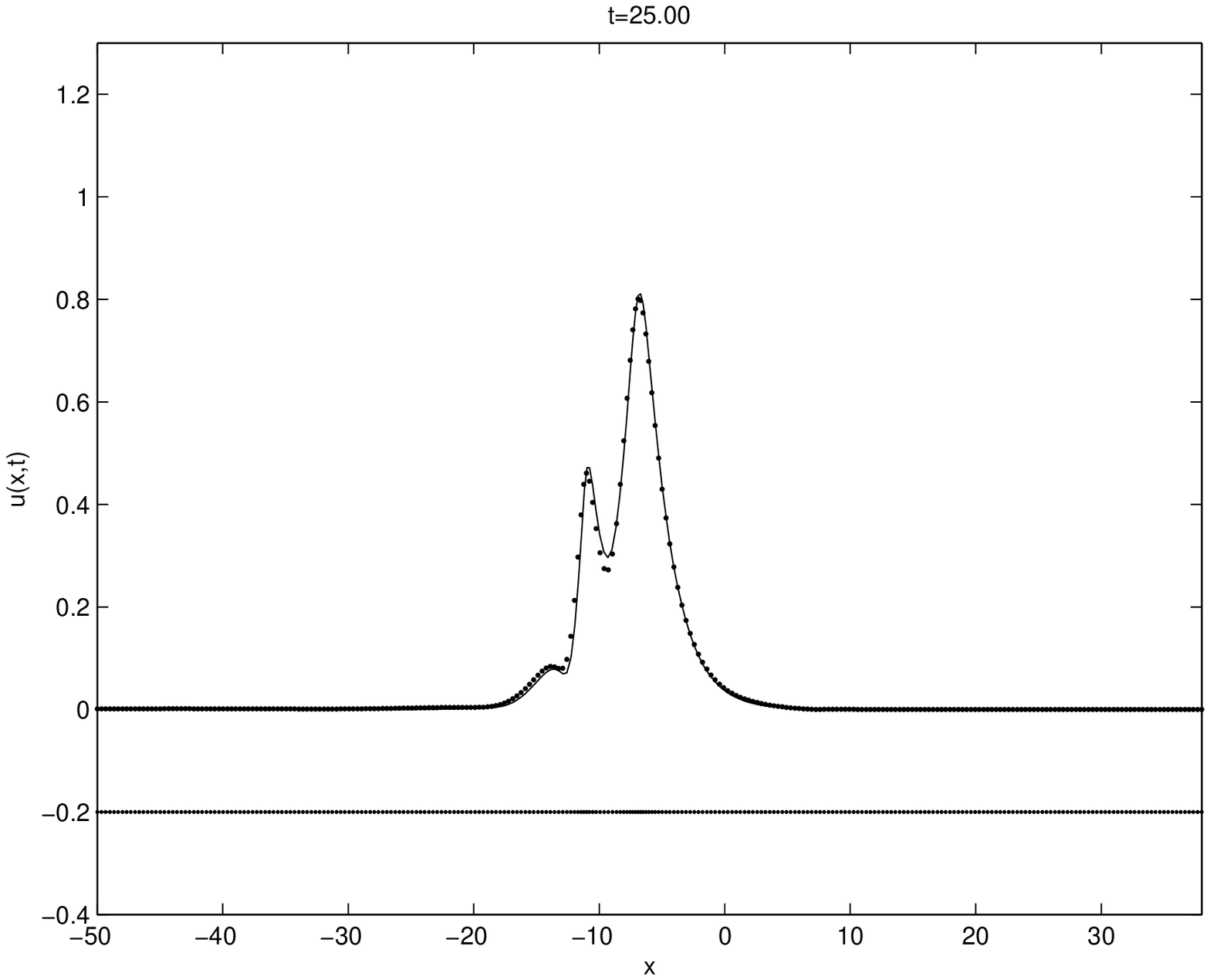} 
\includegraphics[clip,width=6.8cm]{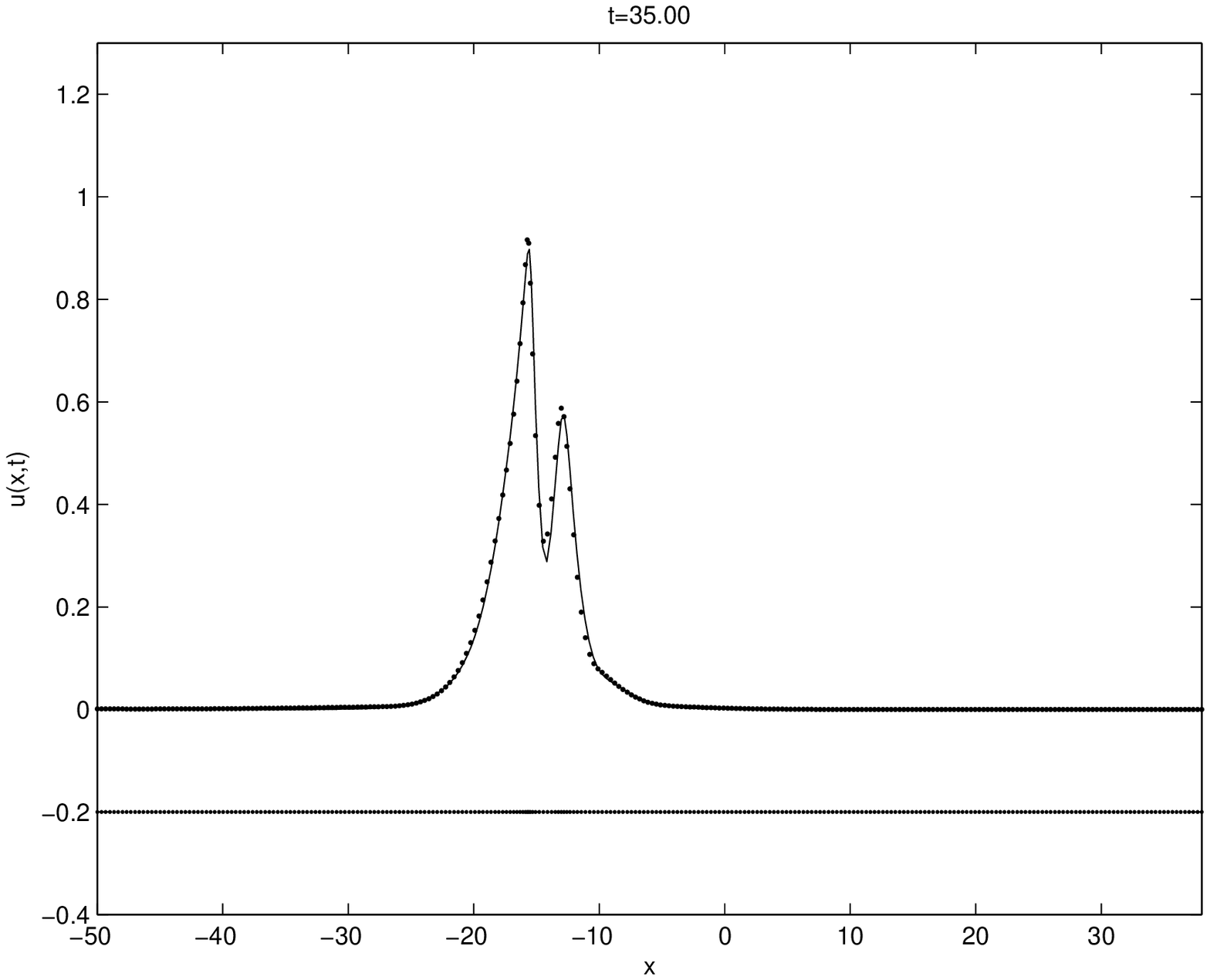} 
\includegraphics[clip,width=6.8cm]{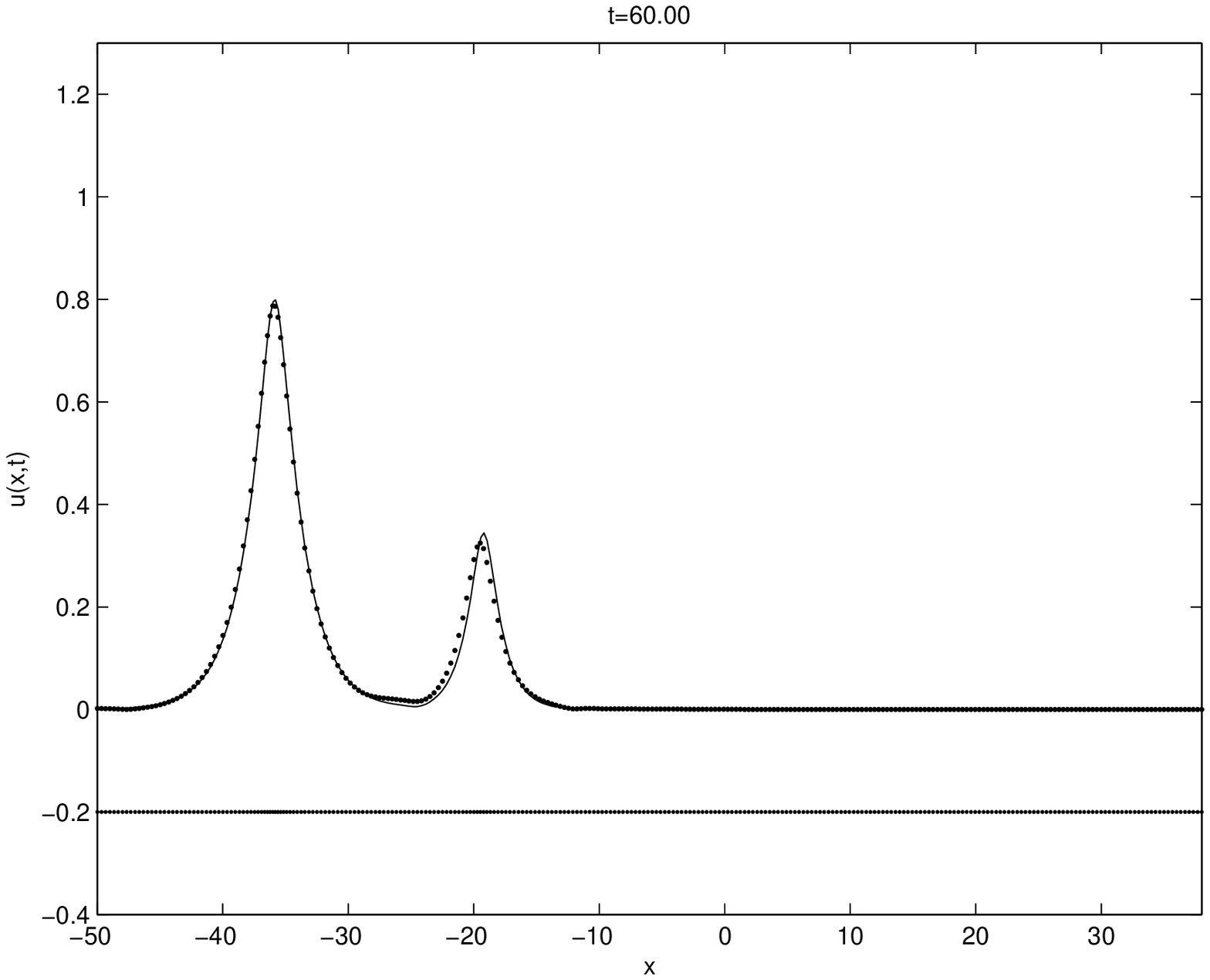} 
\caption{The numerical simulation of the 2 soliton solution of the complex
 SP equation. The points  
shows the numerical values, and the continuous curve shows the exact value, and
 the points on the bottom of graphs show the distribution of mesh grid
 points.  
$p_1=0.5+{\rm i}$, $p_2=0.8+2{\rm i}$, $a_1=e^{-6}$, $a_2=e^{4}$.
}
\label{csp-numerics}
\end{figure}

\begin{figure}
\includegraphics[clip,width=6.8cm]{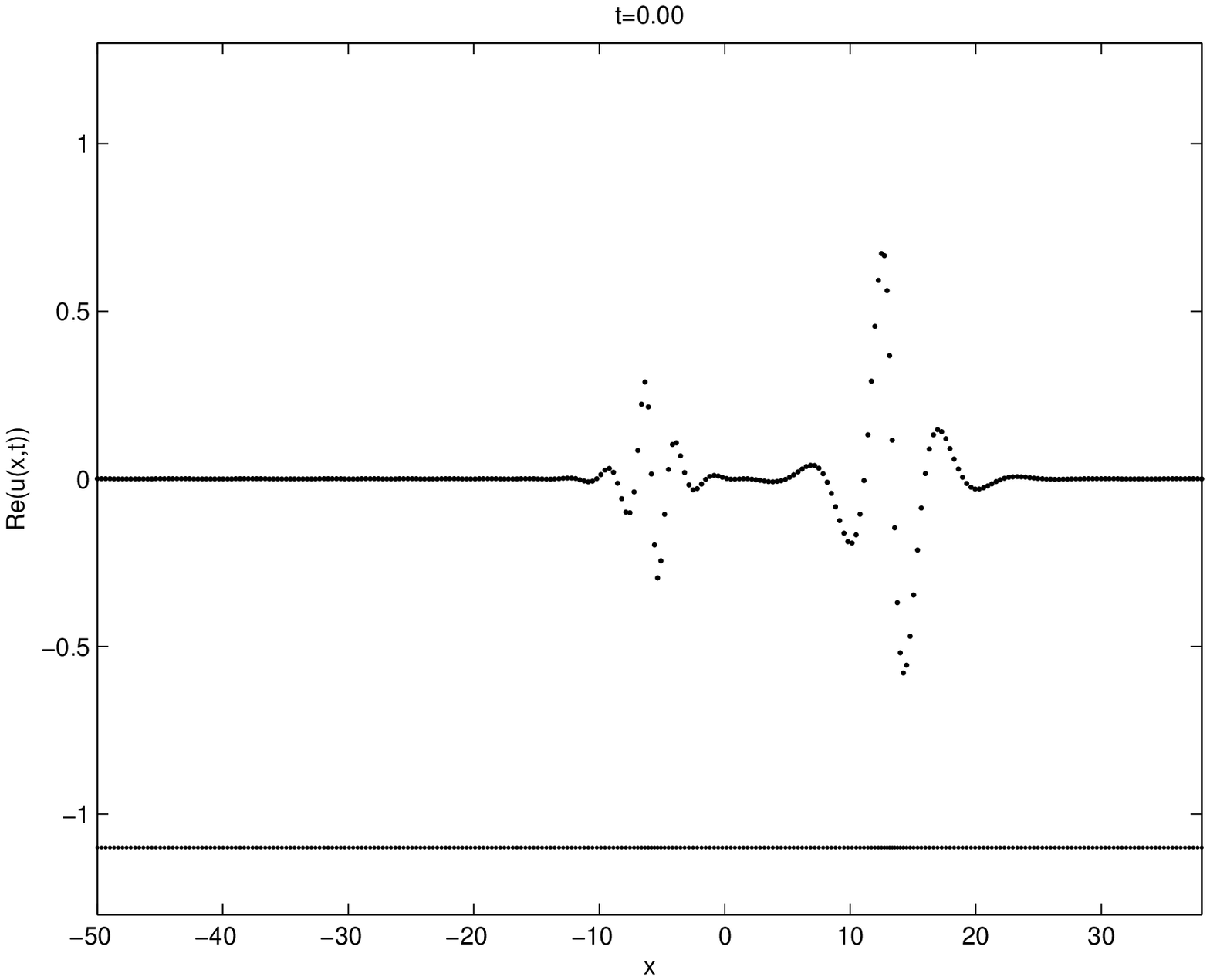} 
\includegraphics[clip,width=6.8cm]{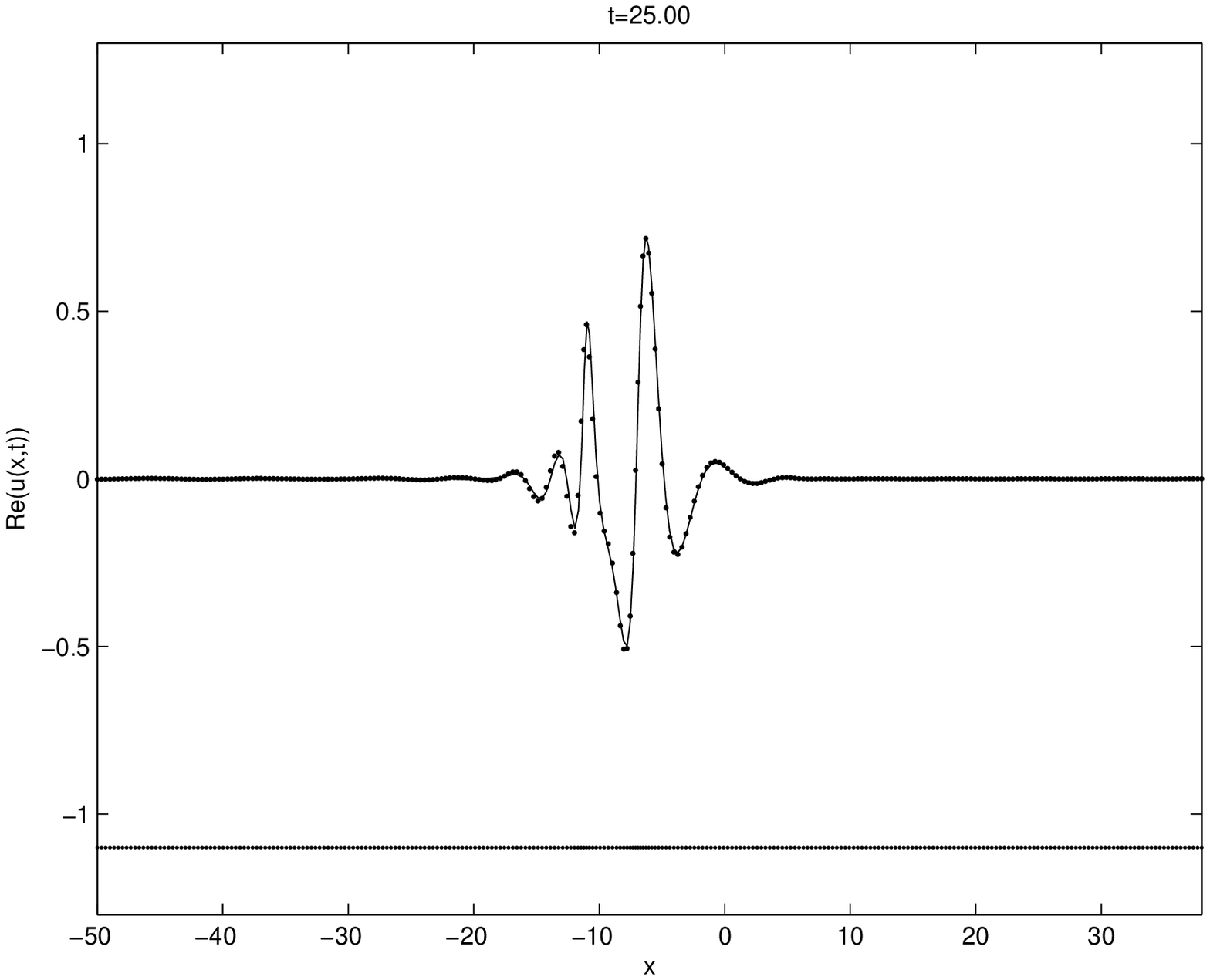} 
\includegraphics[clip,width=6.8cm]{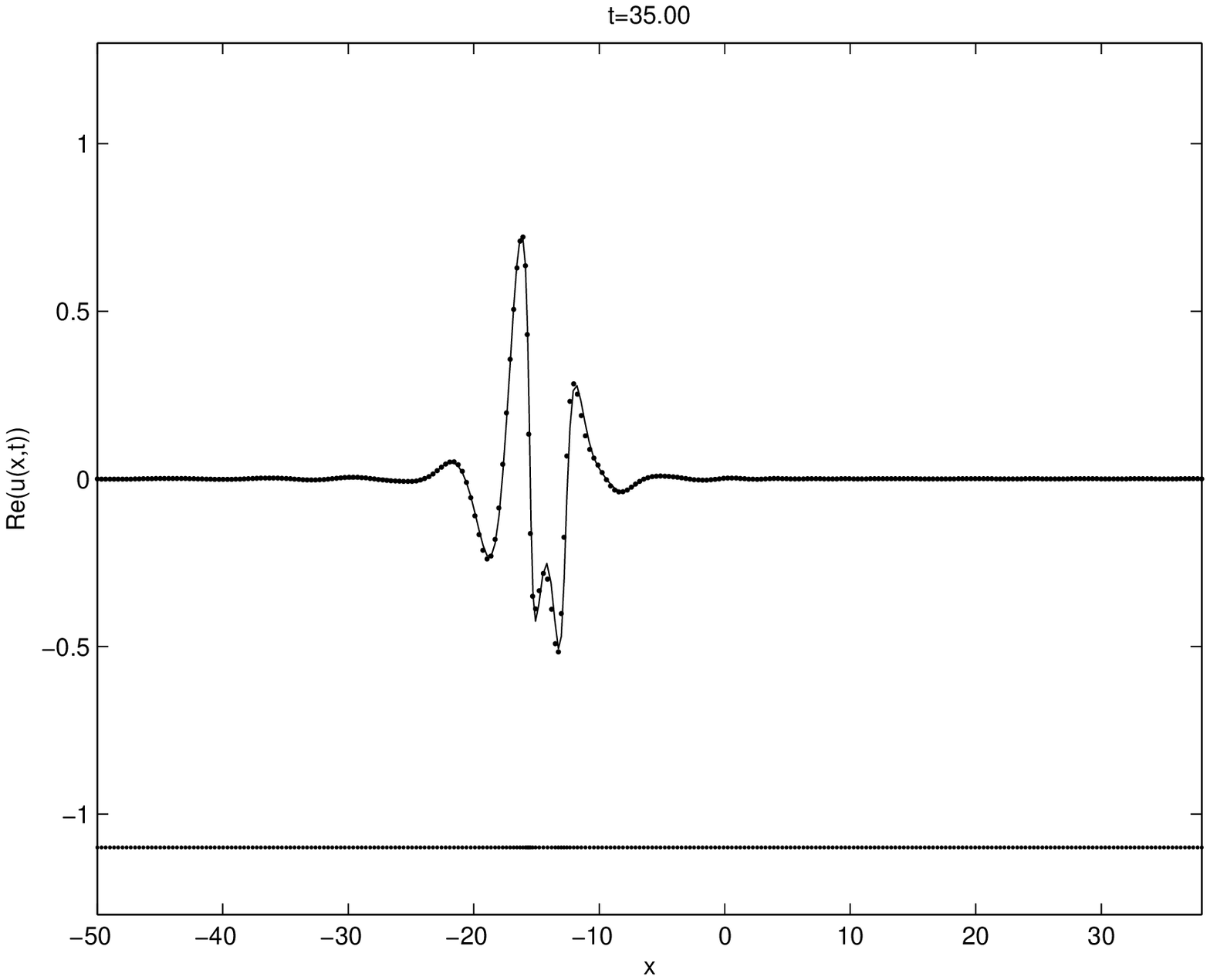} 
\includegraphics[clip,width=6.8cm]{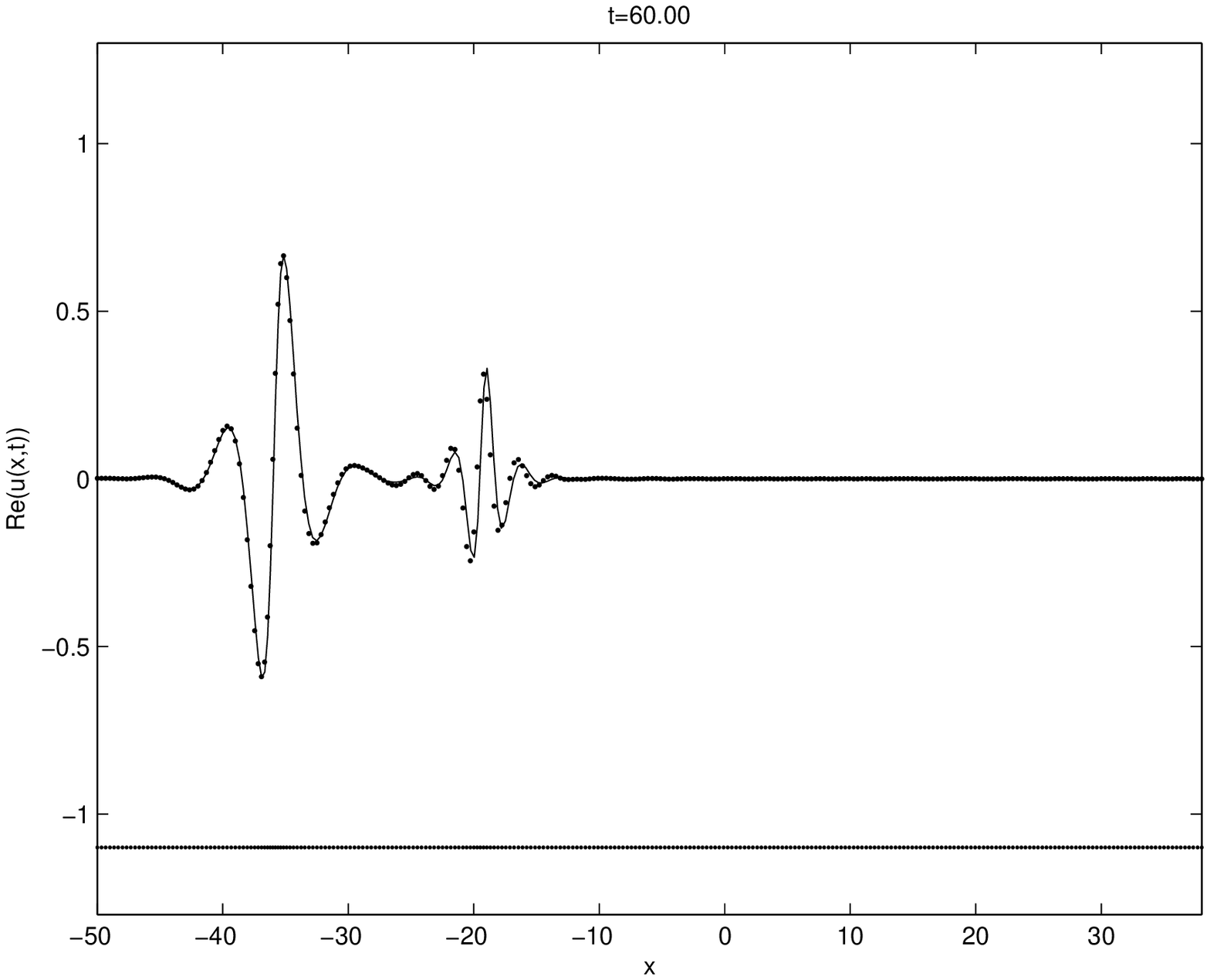} 
\caption{The numerical simulation of the 2 soliton solution of the
 complex SP equation. The graphs show the real part of
 $u$ of the complex SP equation. 
The points shows the numerical values, 
and the continuous curve shows the exact value, and
 the points on the bottom of graphs show the distribution of mesh grid
 points.  
$p_1=0.5+{\rm i}$, $p_2=0.8+2{\rm i}$, $a_1=e^{-6}$, $a_2=e^{4}$.
}
\label{csp2-numerics}
\end{figure}

\begin{figure}
\includegraphics[clip,width=6.8cm]{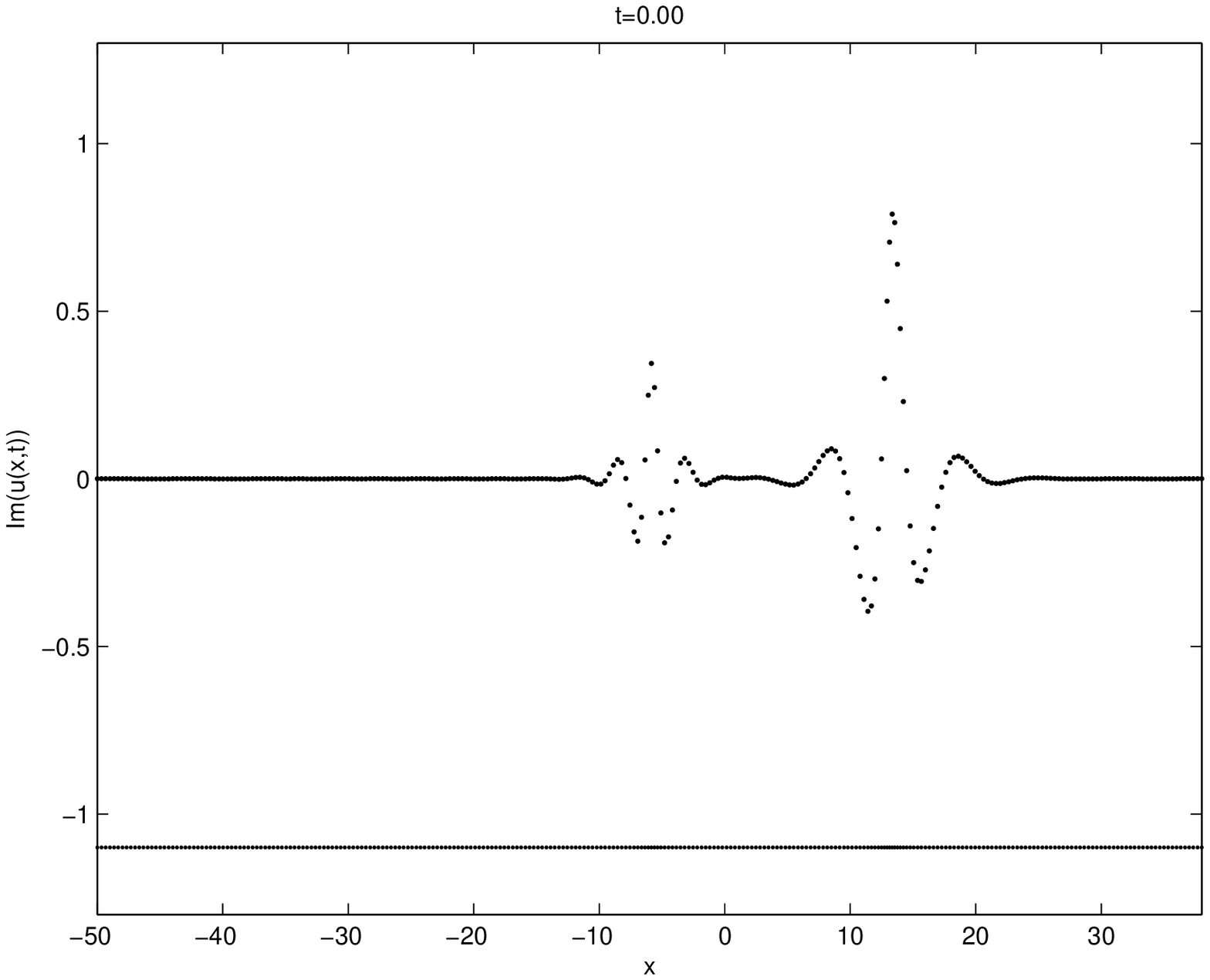} 
\includegraphics[clip,width=6.8cm]{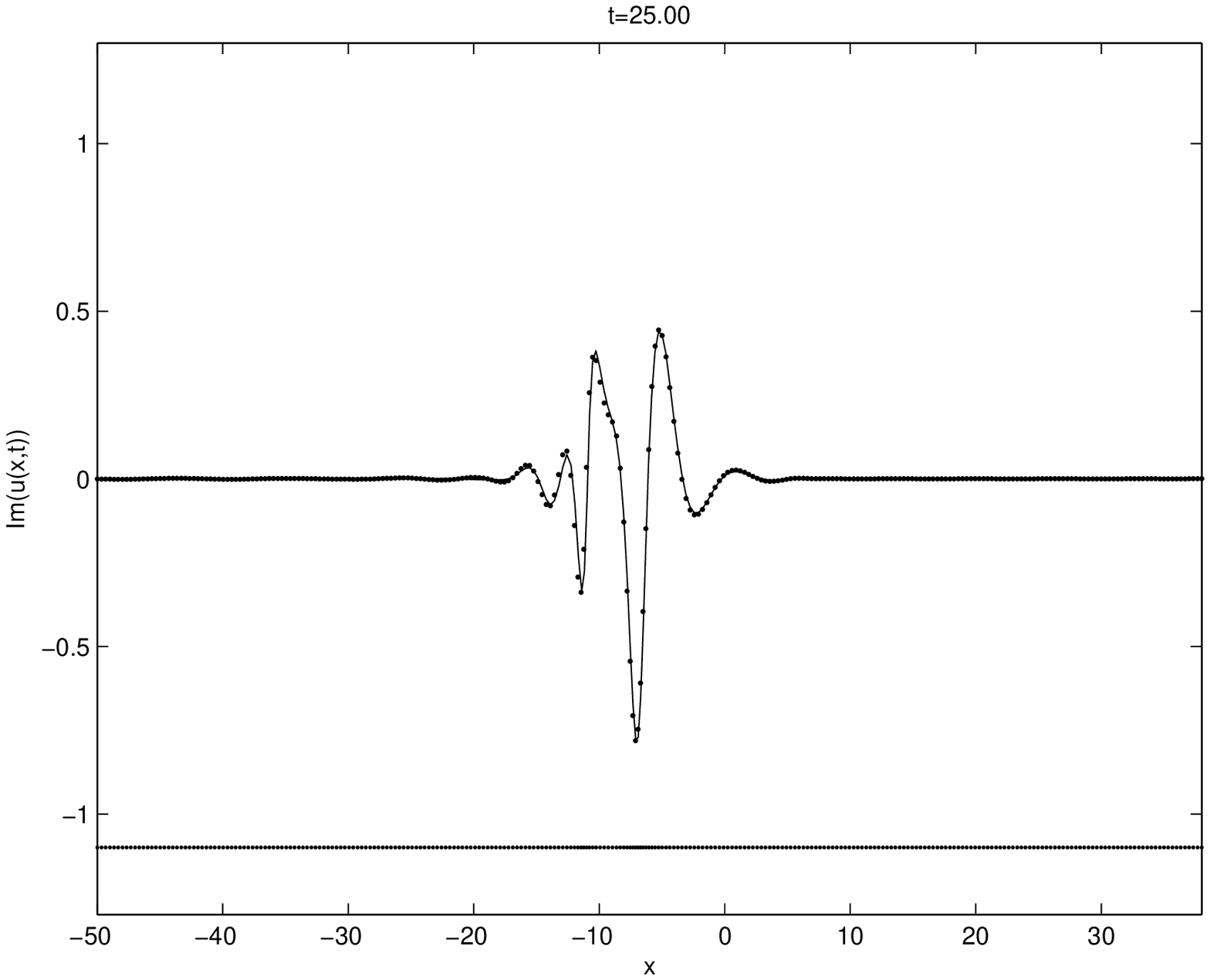} 
\includegraphics[clip,width=6.8cm]{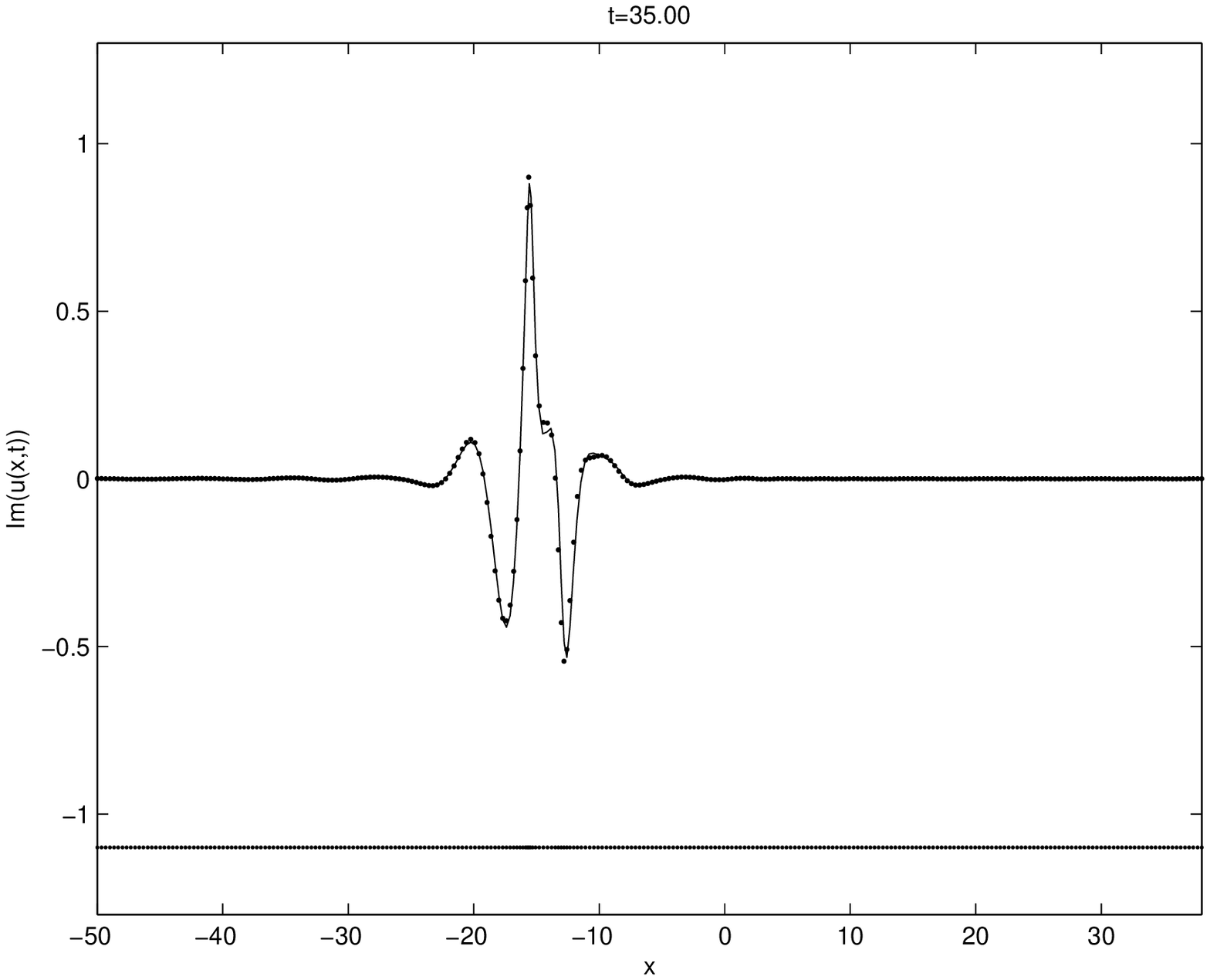} 
\includegraphics[clip,width=6.8cm]{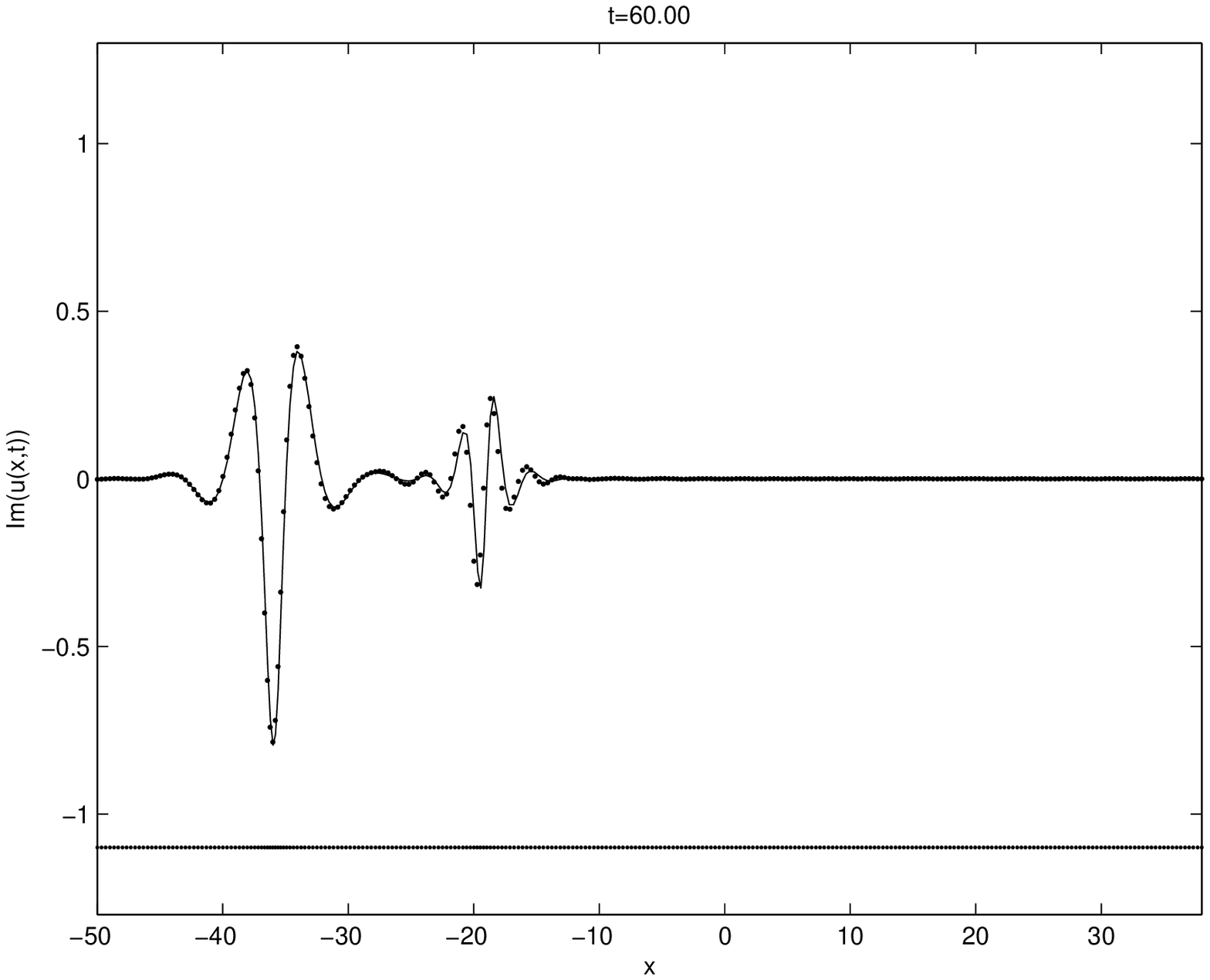} 
\caption{The numerical simulation of the 2 soliton solution of the
 complex SP equation. The graphs show the imaginary part of
 $u$ of the complex SP equation. 
The points shows the numerical values, 
and the continuous curve shows the exact value, and
 the points on the bottom of graphs show the distribution of mesh grid
 points.  
$p_1=0.5+{\rm i}$, $p_2=0.8+2{\rm i}$, $a_1=e^{-6}$, $a_2=e^{4}$.
}
\label{csp3-numerics}
\end{figure}

Figure \ref{csp-numerics} shows the numerical simulation of the
2 soliton solution of the complex SP equation (\ref{complexSPeq1}) and
(\ref{complexSPeq2}) 
by means of the
self-adaptive moving mesh scheme (\ref{semidiscrete-complexSP1}), 
(\ref{semidiscrete-complexSP2}), and (\ref{semidiscrete-complexSP3}). 
Figure \ref{csp2-numerics} and \ref{csp3-numerics} 
show the graphs of the real part and the imaginary part of
$u$ of the complex SP equation (\ref{complexSPeq1}) and
(\ref{complexSPeq2}), respectively. 
Thus these graphs are the solution of 
equations (\ref{complexSPeq1-real}) and (\ref{complexSPeq2-imaginary}). 
We use the number of mesh grid points $N=300$, the width of the computational
domain ${\rm D}=100$, and the time interval $dt=0.00005$.  
Again, the numerical result has good agreement with the exact solution
of the complex SP equation. 


\section{Concluding Remarks}

We have proposed two systematic methods for constructing self-adaptive 
moving mesh schemes for 
a class of nonlinear wave equations which are transformed into 
a different class of nonlinear wave equations by reciprocal (hodograph) 
transformations. 
We have demonstrated how to create self-adaptive moving mesh schemes 
for short pulse type equations which are 
transformed into coupled dispersionless type systems by a reciprocal
(hodograph) transformation. 
Self-adaptive moving mesh schemes have exact solutions such as
multi-soliton solutions and Lax pairs, thus those schemes are
integrable. 
Self-adaptive moving mesh schemes consist of two semi-discrete
equations in which the time is continuous and the space is discrete. 
In self-adaptive moving mesh schemes, one of two equations is an evolution
equation of mesh intervals which is deeply related to 
a discrete analogue of a reciprocal (hodograph) transformation. 
An evolution equations of mesh intervals is 
a discrete analogue of a conservation law of an original equation, 
and a set of mesh intervals corresponds to a conserved density which
play a key role in generation of adaptive moving mesh. 
We have shown several examples of numerical computations of 
the short pulse type equations by using
self-adaptive moving mesh schemes. 

In our previous papers, we have investigated how to discretize the
Camassa-Holm\cite{dCH,dCHcom}, the Hunter-Saxton\cite{dHS}, 
the short pulse\cite{SPE_discrete1,SPE_discrete2}, the WKI elastic beam\cite{SPE_discrete2}, 
the Dym equation\cite{SPE_discrete1,dDym} 
by using bilinear methods or by using a geometric approach. 
Based on our previous studies, we have 
proposed two systematic methods in sophisticated forms, 
one uses bilinear equations and another uses Lax pairs, for producing 
self-adaptive moving mesh schemes. 
Although we have discussed only short pulse type equations in this paper, 
our methods can be used to construct self-adaptive moving mesh schemes 
for other nonlinear wave equations in the WKI class. 

More details about exact solutions, fully discretizations, and numerical computations 
of self-adaptive moving mesh 
schemes for the coupled SP equation, the complex SP equation, and their
generalized equations will be discussed in our forthcoming
papers. 

\section*{Appendix}

The multi-soliton solutions of the coupled SP equation
(\ref{coupledSPeq1}) and (\ref{coupledSPeq2}) are given by the
following formula:
\begin{eqnarray}
&&u=\frac{g}{f}\,, \quad  v=\frac{h}{f}\,, \quad \rho=1-2(\ln f)_{XT}\,,\\
&&x=X_{0}+\int_{X_{0}}^{X}\rho(\tilde{X},T) d\tilde{X} \,, \quad
 t=T\,,\\
&& f=\left|
\begin{matrix}
\mathcal{A}_N &I_N \\
\noalign{\vskip5pt} -I_N & \mathcal{B}_N
\end{matrix}
\right|
=\left|I_N+\mathcal{A}_N\mathcal{B}_N\right|
\,,\\
&& g={\left|
\begin{matrix}
\mathcal{A}_N &I_N & {{\bf e}_N^{(1)}}^\top\\
\noalign{\vskip5pt}
-I_N & \mathcal{B}_N & {\bf 0}^\top\\
\noalign{\vskip5pt}
{\bf 0} & -{\bf a}_N^{(1)} & 0
\end{matrix}
\right|}\,,\\
&& h={\left|
\begin{matrix}
\mathcal{A}_N &I_N & {\bf 0}^\top\\
\noalign{\vskip5pt}
-I_N & \mathcal{B}_N & {{\bf a}_N^{(2)}}^\top\\
\noalign{\vskip5pt}
{\bf e}_N^{(2)} & {\bf 0} & 0
\end{matrix}
\right|}\,,
\end{eqnarray}
where 
\begin{eqnarray*}
&&\mathcal{A}_N=\\
&&\small
\left(
\begin{matrix}
\frac{{\rm e}^{\xi_1+\xi_{N+1}}}{4\left(\frac{1}{p_1}+\frac{1}{p_{N+1}}\right)}
&\frac{{\rm e}^{\xi_1+\xi_{N+2}}}{4\left(\frac{1}{p_1}+\frac{1}{p_{N+2}}\right)}
&\cdots
&\frac{{\rm e}^{\xi_1+\xi_{2N}}}{4\left(\frac{1}{p_1}+\frac{1}{p_{2N}}\right)}
\\
\frac{{\rm e}^{\xi_2+\xi_{N+1}}}{4\left(\frac{1}{p_2}+\frac{1}{p_{N+1}}\right)}
&\frac{{\rm e}^{\xi_2+\xi_{N+2}}}{4\left(\frac{1}{p_2}+\frac{1}{p_{N+2}}\right)}
&\cdots
&\frac{{\rm e}^{\xi_2+\xi_{2N}}}{4\left(\frac{1}{p_2}+\frac{1}{p_{2N}}\right)}
\\
\vdots&\vdots&\ddots&\vdots
\\
\frac{{\rm e}^{\xi_N+\xi_{N+1}}}{4\left(\frac{1}{p_N}+\frac{1}{p_{N+1}}\right)}
&\frac{{\rm e}^{\xi_N+\xi_{N+2}}}{4\left(\frac{1}{p_N}+\frac{1}{p_{N+2}}\right)}
&\cdots
&\frac{{\rm e}^{\xi_N+\xi_{2N}}}{4\left(\frac{1}{p_N}+\frac{1}{p_{2N}}\right)}
\end{matrix}
\right)
\normalsize
\,,
\end{eqnarray*}
\[
\mathcal{B}_N= 
\left(
\begin{matrix}
\frac{a_1a_{N+1}}{1/p_1+1/p_{N+1}}
&\frac{a_2a_{N+1}}{1/p_2+1/p_{N+1}}&\cdots
&\frac{a_Na_{N+1}}{1/p_N+1/p_{N+1}}
\\
\frac{a_1a_{N+2}}{1/p_1+1/p_{N+2}}
&\frac{a_2a_{N+2}}{1/p_2+1/p_{N+2}}
&\cdots
&\frac{a_Na_{N+2}}{1/p_N+1/p_{N+2}}
\\
\vdots&\vdots&\ddots&\vdots
\\
\frac{a_1a_{2N}}{1/p_1+1/p_{2N}}
&\frac{a_2a_{2N}}{1/p_2+1/p_{2N}}&\cdots
&\frac{a_Na_{2N}}{1/p_N+1/p_{2N}}
\end{matrix}
\right)\,,
\]
and $I_N$ is the $N\times N$ identity matrix, 
${\bf a}^\top$ is the transpose of ${\bf a}$, 
\begin{eqnarray*}
&&{\bf a}_N^{(1)}=(a_1,a_2,\cdots ,a_N)\,,\quad 
{\bf e}_N^{(1)}=(e^{\xi_1},e^{\xi_2},\cdots ,e^{\xi_N}) \,,\\ 
&&{\bf a}_N^{(2)}=(a_{N+1},a_{N+2},\cdots ,a_{2N})\,,\\
&&{\bf e}_N^{(2)}=(e^{\xi_{N+1}},e^{\xi_{N+2}},\cdots ,e^{\xi_{2N}}) \,,\\ 
&& {\bf 0}=(0,0,\cdots , 0)\,, 
\end{eqnarray*}
\[
\xi_i=p_iX+\frac{1}{p_i}T\,,
\qquad 1\le i\le 2N\,. 
\]



\begin{thebibliography}{99}

\bibitem{Hirota:difference1} 
Hirota, R.:
Nonlinear partial difference equations. I. 
A difference analogue of the Korteweg-de Vries equation, 
{\em J. Phys. Soc. Jpn.} \textbf{43} (1977) 4116--4124.

\bibitem{Hirota:difference2} 
Hirota, R.:
Nonlinear partial difference equations. II. Discrete-time Toda equation, 
{\em J. Phys. Soc. Jpn.} \textbf{43} (1977) 2074--2078.

\bibitem{Hirota:difference3} 
Hirota, R.:
Nonlinear partial difference equations. III. Discrete sine-Gordon equation, 
{\em J. Phys. Soc. Jpn.} \textbf{43} (1977)  2079--2086.

\bibitem{Hirota:difference4} 
Hirota, R.: 
Nonlinear partial difference equations. IV. B\"acklund
transformation for the discrete-time Toda equation, 
{\em J. Phys. Soc. Jpn.} \textbf{45} (1978) 321--332.

\bibitem{Hirota:difference5} 
Hirota, R.:
Nonlinear partial difference equations. V. 
Nonlinear equations reducible to linear equations,
{\em J. Phys. Soc. Jpn.} \textbf{46} (1979) 312--319.

\bibitem{Ablowitz-Ladik:1975} 
Ablowitz, M. J. and Ladik, J. F.:
Nonlinear differential-difference equations, 
{\em J. Math. Phys.} \textbf{16} (1975) 598--603.

\bibitem{Ablowitz-Ladik:1976} 
Ablowitz, M. J. and Ladik, J. F.: 
Nonlinear differential-difference equations and Fourier analysis, 
{\em J. Math. Phys.} \textbf{17} (1976) 1011--1018.

\bibitem{Ablowitz-Ladik:1976-2} 
Ablowitz, M. J. and Ladik, J. F.: 
A nonlinear difference scheme and inverse scattering, 
{\em Stud. Appl. Math.} \textbf{55} (1977) 213--229.

\bibitem{Ablowitz-Ladik:1977} 
Ablowitz, M. J. and Ladik, J. F.: 
On the solution of a class of nonlinear partial difference equations, 
{\em Stud. Appl. Math.} \textbf{57} (1977) 1--12.

\bibitem{AS:book} 
Ablowitz, M. J. and Segur, H.:
{\em Solitons and Inverse Scattering Transform}
SIAM, Philadelphia, 1981.

\bibitem{Grammaticos:book} 
Grammaticos, B., Kosmann-Schwarzbach, Y. and Tamizhmani, T. (Eds):
{\em Discrete Integrable Systems}, 
Lecture Notes in Physics \textbf{644}, Springer-Verlag, Berlin, 2004.

\bibitem{Suris:book} 
Suris, Y. B.:
{\em The Problem of Integrable Discretization: Hamiltonian Approach}
Birkh\"auser, Basel, 2003.

\bibitem{Bobenko:book}
Bobenko, A. I. and Suris, Y. B.:
{\em Discrete Differential Geometry},
Graduate Studies in Mathematics \textbf{98},
AMS, Rhode Island, 2008.

\bibitem{Bobenko:book2}
Bobenko, A. I. and Seiler, R. (Eds.):
{\em Discrete Integrable Geometry and Physics},
Oxford Lecture Series in Mathematics and Its Applications
	\textbf{16}, 
Oxford Univ. Press, Oxford, 1999.

\bibitem{WKI1}
Wadati, M., Konno, K. and Ichikawa, Y.:
New integrable nonlinear evolution equations, 
{\em J. Phys. Soc. Jpn.} \textbf{47} (1979) 1698--1700. 

\bibitem{WKI2}
Konno, K., Ichikawa, Y. and Wadati, M.:
A loop soliton propagating along a stretched rope,
{\em J. Phys. Soc. Jpn.} \textbf{50} (1981) 1025--1026.

\bibitem{Ishimori1}
Ishimori, Y.: 
On the modified Korteweg-de Vries soliton and the loop soliton, 
{\em J. Phys. Soc. Jpn.} \textbf{50} (1981) 2471--2472.

\bibitem{Ishimori2}
Ishimori, Y.: 
A relationship between the Ablowitz-Kaup-Newell-Segur 
and Wadati-Konno-Ichikawa schemes of the inverse scattering method, 
{\em J. Phys. Soc. Jpn.} \textbf{51} (1982) 3036--3041.

\bibitem{Wadati-Sogo}
Wadati, S. and Sogo, K.: 
Gauge transformations in soliton theory, 
{\em J. Phys. Soc. Jpn.} \textbf{52} (1983) 394--398.

\bibitem{Rogers-Schief}
Rogers, S. and Schief, W. K.: 
{\em B\"acklund and Darboux Transformations: Geometry and Modern Applications in Soliton Theory},
Cambridge University Press, Cambridge, 2002.

\bibitem{dCH}
Ohta, Y.,  Maruno, K. and Feng, B. F.: 
An integrable semi-discretization of the 
Camassa-Holm equation and its determinant solution, 
{\em J. Phys. A} \textbf{41} (2008) 355205.

\bibitem{dCHcom}
Feng, B. F, Maruno, K. and Ohta, Y.: 
A self-adaptive moving mesh method for the Camassa-Holm equation, 
{\em J. Comput. Appl. Math.} \textbf{235} (2010) 229--243.

\bibitem{dHS}
Ohta, Y.,  Maruno, K. and Feng, B. F.: 
Integrable discretizations for the short-wave 
model of the Camassa–Holm equation, 
{\em J. Phys. A} \textbf{43} (2010) 265202.

\bibitem{SPE_discrete1} 
Feng, B. F., Maruno, K. and Ohta, Y.:
Integrable discretizations of the short pulse equation, 
{\em J. Phys. A} \textbf{43} (2010) 085203.

\bibitem{SPE_discrete2} 
Feng, B. F., Inoguchi, J., Kajiwara, K., Maruno, K. and Ohta, Y.: 
Discrete integrable systems and hodograph transformations arising 
from motions of discrete plane curves,
{\em J. Phys. A} \textbf{44} (2011) 395201.

\bibitem{dDym}
Feng, B. F., Inoguchi, J., Kajiwara, K., Maruno, K. and Ohta, Y.: 
Integrable discretizations of the Dym equation, 
{\em Front. Math. in China}, 
\textbf{8} (2013) 1017--1029

\bibitem{SP:SW}
Sch\"afer, T. and Wayne, C. E.: 
Propagation of ultra-short optical pulses in cubic nonlinear media,
{\em Physica D} \textbf{196} (2004) 90--105.

\bibitem{SP:CJSW}
Chung, Y., Jones, C. K. R. T., Sch\"afer, T. and Wayne, C. E.: 
Ultra-short pulses in linear and nonlinear media,
{\em Nonlinearity} \textbf{18} (2005) 1351--1374.

\bibitem{Sakovich1}
Sakovich, A. and Sakovich, S.: 
The short pulse equation is integrable, 
{\em J. Phys. Soc. Jpn.} \textbf{74} (2005) 239--241.

\bibitem{Sakovich2}
Sakovich, A. and Sakovich, S.: 
Solitary wave solutions of the short pulse equation, 
{\em J. Phys. A} \textbf{39} (2006) L361--367.

\bibitem{Matsuno_SPE}
Matsuno, Y.: 
Multiloop soliton and multibreather solutions of 
the short pulse model equation, 
{\em J. Phys. Soc. Jpn.} \textbf{76} (2007) 084003.

\bibitem{Matsuno_SPEreview}
Matsuno, Y.: 
Soliton and periodic solutions of the short pulse model equation, 
In: Lang, S. P. and Bedore, H. (Eds.) 
{\em Handbook of Solitons: Research, Technology and Applications}, 
(2009) 541--586, Nova Publishers.

\bibitem{Konno-Oono1}
Konno, K. and Oono, H.: 
New coupled integrable dispersionless equations, 
{\em J. Phys. Soc. Jpn.} \textbf{63} (1994) 377--378.

\bibitem{Hirota-Tsujimoto}
Hirota, R. and Tsujimoto, S.: 
Note on ``New coupled integrable dispersionless equations'',
{\em J. Phys. Soc. Jpn.} \textbf{63} (1994) 3533.

\bibitem{Konno-Oono2}
Konno, K. and Oono, H.: 
Reply to note on ``New coupled integrable dispersionless equations'', 
{\em J. Phys. Soc. Jpn.} \textbf{63} (1994) 3534.

\bibitem{Konno}
Konno, K.: 
Integrable coupled dispersionless equations, 
{\em Applicable Analysis} \textbf{57} (1995) 209--220.

\bibitem{Kakuhata-Konno1}
Kakuhata, H. and Konno, K.: 
Interaction among growing, decaying and stationary solitons for coupled
	integrable 
dispersionless equations, 
{\em J. Phys. Soc. Jpn.} \textbf{64} (1995) 2707--2709.

\bibitem{Kakuhata-Konno2}
Kakuhata, H. and Konno, K.: 
Novel solitonic evolutions in a coupled integrable, dispersionless
	system, 
{\em J. Phys. Soc. Jpn.} \textbf{65} (1996) 340--341.


\bibitem{Matsuno-CSPE}
Matsuno, Y.: 
A novel multi-component generalization of the short pulse equation and its multisoliton solutions, 
{\em J. Math. Phys.} \textbf{52} (2011) 123702.

\bibitem{Folkert-CSPE}
Dimakis, A. and M\"uller-Hoissen, F.: 
Bidifferential calculus approach to AKNS hierarchies and their
	solutions, 
{\em SIGMA} \textbf{6} (2010) 055.

\bibitem{Kotlyarov}
Kotlyarov, V.: 
On equations gauge equivalent to the sine-Gordon and
	Pohlmeyer-Lund-Regge equations,
{\em J. Phys. Soc. Jpn.} \textbf{63} (1994) 3535--3537. 

\bibitem{Feng-CSPE} 
Feng, B. F.:
Complex short pulse and coupled complex short pulse equations, 
{\em arXiv:1312.6431} (2013).

\bibitem{Vinet-Yu1}
Vinet, L. and Yu, G-F.: 
On the discretization of the coupled integrable dispersionless
	equations, 
{\em J. Nonlinear Math. Phys.} \textbf{20} (2013) 106--125.

\bibitem{Vinet-Yu2}
Vinet, L. and Yu, G-F.: 
Discrete analogues of the generalized coupled integrable 
dispersionless equations, 
{\em J. Phys. A}  \textbf{46} (2013) 175205.

\end{thebibliography}
\end{document}